\documentclass[manuscript,screen,nonacm]{acmart}
\setcopyright{none}
\usepackage{algorithmic}
\usepackage{algorithm}
\usepackage{wrapfig}
\usepackage{multirow}
\usepackage{makecell}
\usepackage{threeparttable}
\usepackage{tabularx}
\usepackage{tikz}
\usepackage{pifont}
\begin{document}
\raggedbottom
\title[Structure-Aligned History Learning for Generative Recommendation]{Structure-Aligned History Learning with Subspace-Structured Semantic IDs for Generative Recommendation}
\author{Kaiwen Wei}
\authornote{Kaiwen Wei and Kejun He contributed equally to this work.}
\affiliation{\institution{College of Computer Science, Chongqing University}\city{Chongqing}\country{China}}
\email{weikaiwen@cqu.edu.cn}
\author{Kejun He}
\authornotemark[1]
\affiliation{\institution{College of Computer Science, Chongqing University}\city{Chongqing}\country{China}}
\email{hekejun@stu.cqu.edu.cn}
\author{Jie Zhang}
\affiliation{\institution{China Telecom Artificial Intelligence Technology (Beijing) Co. Ltd}\city{Beijing}\country{China}}
\email{zhangj157@chinatelecom.cn}
\author{Xiaomian Kang}
\affiliation{\institution{MAIS, Institute of Automation, Chinese Academy of Sciences}\city{Beijing}\country{China}}
\email{xiaomian.kang@nlpr.ia.ac.cn}
\author{Yuming Yang}
\affiliation{\institution{College of Computer Science, Chongqing University}\city{Chongqing}\country{China}}
\email{ymyang@cqu.edu.cn}
\author{Aijia Zhou}
\affiliation{\institution{College of Computer Science, Chongqing University}\city{Chongqing}\country{China}}
\email{202414131135@stu.cqu.edu.cn}
\author{Xie Zhou}
\affiliation{\institution{University of Science and Technology Beijing}\city{Beijing}\country{China}}
\email{zhouxie@xs.ustb.edu.cn}
\author{Yining Wang}
\affiliation{\institution{Unisound AI}\city{Beijing}\country{China}}
\email{wangyining@unisound.com}
\author{Nayu Liu}
\affiliation{\institution{School of Computer Science and Technology, Tianjin University}\city{Tianjin}\country{China}}
\email{nyliu@tju.edu.cn}
\author{Li Jin}
\affiliation{\institution{Minzu University of China}\city{Beijing}\country{China}}
\email{jinlimails@gmail.com}
\author{Zhenyang Li}
\affiliation{\institution{The Hong Kong Polytechnic University}\city{Hong Kong SAR}\country{China}}
\email{zhenyanglidz@gmail.com}
\author{Jiang Zhong}
\authornote{Corresponding authors: Jiang Zhong and Junnan Zhu.}
\affiliation{\institution{College of Computer Science, Chongqing University}\city{Chongqing}\country{China}}
\email{zhongjiang@cqu.edu.cn}
\author{He Bai}
\affiliation{\institution{Independent researcher}\country{China}}
\email{richardbai@apple.com}
\author{Jingxuan Wei}
\affiliation{\institution{Shandong Computer Science Center (National Supercomputer Center in Jinan)}\city{Jinan}\country{China}}
\email{weijx@sdas.org}
\author{Junnan Zhu}
\authornotemark[2]
\affiliation{\institution{MAIS, Institute of Automation, Chinese Academy of Sciences}\city{Beijing}\country{China}}
\email{junnan.zhu@nlpr.ia.ac.cn}
\renewcommand{\shortauthors}{Wei et al.}
\begin{abstract}

Generative recommendation constructs semantic IDs for catalog items and predicts the next ID from user interaction histories. However, existing methods face limitations in both stages. First, conventional quantizers may produce imbalanced code assignments and lack coherent intermediate semantic units between individual tokens and complete items, limiting history reconstruction. Second, recommendation models remain dominated by next-item prediction, treating historical interactions mainly as conditioning context and providing limited supervision for recovering persistent preferences across the historical path, particularly when relevant preference signals are distributed over distant interactions.
To address these limitations, Structure-Aligned History Learning (SAHL) reconnects the two stages by using the structure created during quantization directly for history reconstruction. Its Semantic Residual Balanced Quantization (SRBQ) forms a balanced token--subspace--item hierarchy, and SAHL reconstructs history at these same granularities. Predictive entropy prioritizes uncertain transitions, while curriculum optimization progressively shifts training to inference-aligned next-item prediction. On four Amazon Review categories, SAHL improves NDCG@10 by 18.6\%--30.0\% over RPG and 2.7\%--6.0\% over strong method MHL. Controlled analyses verify balanced code use, the benefit of following quantizer-defined groups during reconstruction, and dependence on user history. The relevant code is available at \url{https://github.com/CQU-MM-Intelligent-Lab/SAHL}.

\end{abstract}
\maketitle
\section{Introduction}

\begin{wrapfigure}{R}{0.5\textwidth}
 \centering
 \includegraphics[width=\linewidth]{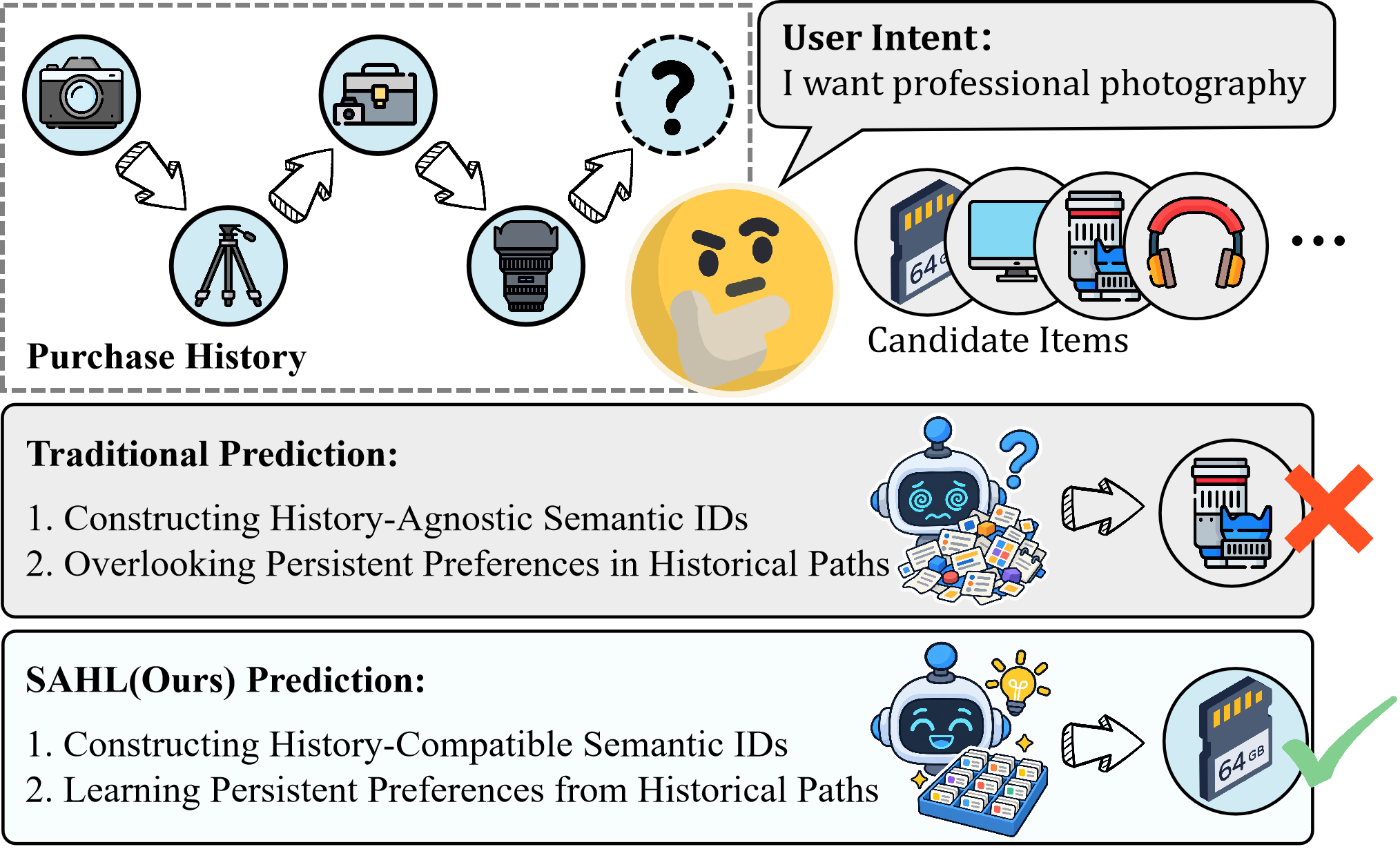}
 \caption{Motivation for the proposed SAHL method. SAHL aligns hierarchical semantic IDs with multi-granularity history reconstruction to capture persistent interests beyond short-range item co-occurrences.}
 \label{fig:intro}
\Description{Comparison of semantic-ID construction and history-learning paradigms, motivating aligned reconstruction units.}
\end{wrapfigure}

Recommender systems have evolved from collaborative filtering~\cite{wang2015collaborative,li2024surveydeepneuralnetworks,chen2018survey} to sequential models that capture the dynamics of user behavior~\cite{purificato2024usermodelinguserprofiling,yuan2023where,he2023survey}. More recently, generative recommendation~\cite{rajput2023recommender,wang2024eager,tan2025imbgen,hou2025generating} has emerged as a promising paradigm that leverages pretrained language models~\cite{rajput2023recommender,bao2023tallrec,hou2025generative} to model interaction sequences and generate semantic IDs of recommended items~\cite{hua2023how,wang2025gnpr,zhai2024actions}.

Generative recommendation typically first converts catalog items into semantic IDs and then predicts the next ID from a user's interaction sequence. These two stages jointly determine how effectively the model can learn from historical behavior. Semantic-ID construction defines \emph{what units of an item are available for reconstruction}, while the training objective determines \emph{how these units are exploited to capture dependencies along the interaction history}. In representative generative recommenders~\cite{rajput2023recommender,wang2024eager,tan2025imbgen,hou2025generating}, however, semantic-ID construction and recommendation learning often pursue different primary objectives, which may weaken the representation structure's support for history supervision.

\begin{figure}[!t]
  \centering
  \begin{minipage}{0.5\textwidth}
  \centering
  \includegraphics[width=\linewidth]{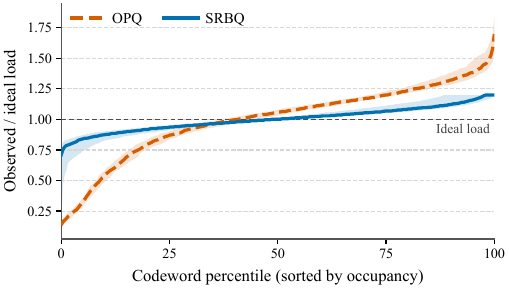}
  \caption{Codebook utilization on the CDs and Vinyl dataset. Codeword occupancies are sorted and normalized by the ideal load. At each occupancy percentile, the curves and bands show the median and 10th--90th percentiles across 64 token positions. OPQ exhibits under-utilized and overloaded codewords, whereas the proposed SRBQ provides more evenly supported reconstruction targets.}
  \label{fig:codebook_profile}
\Description{Codeword utilization profiles for OPQ and SRBQ on CDs and Vinyl, comparing occupancy and peak loads.}
  \end{minipage}
\end{figure}

This mismatch gives rise to two coupled challenges. \textbf{\mbox{(1) Constructing History-Compatible} Semantic IDs.} Optimized product quantization (OPQ)~\cite{jegou2011pq,ge2013optimized} prioritizes compact encoding and distortion reduction, exposing individual tokens and the complete ID but no intermediate units aligned with coherent latent factors; multi-token reconstruction must therefore impose arbitrary groups. Figure~\ref{fig:codebook_profile} also shows uneven OPQ occupancy on the CDs and Vinyl dataset, yielding unevenly supported reconstruction targets. \textbf{(2) Learning \mbox{Persistent Preferences} from Historical Paths.} Conventional next-item objectives~\cite{rajput2023recommender,hou2025generating} primarily use the semantic ID as the prediction target. Although they can learn dependencies from interaction sequences, they do not explicitly require the model to reconstruct preference evidence distributed across earlier interactions. The model may consequently favor recent co-occurrences over persistent preferences. As illustrated in Fig.~\ref{fig:intro}, \mbox{the target \textit{memory card}} is consistent with the preference signaled by the earlier \textit{camera body}, whereas a model focused on the latest \textit{camera lens} may generate lens accessories. Taken together, the representation provides no principled intermediate units for history reconstruction, while the objective provides no direct supervision for learning from such units along the historical path.

To restore this connection, we propose \textbf{Structure-Aligned History Learning} (SAHL). Its \textbf{Semantic Residual Balanced Quantization} (SRBQ) partitions decorrelated dimensions into subspaces, quantizes residuals within each group, and controls occupancy. This process establishes a balanced \emph{token} $\subset$ \emph{subspace} $\subset$ \emph{item} hierarchy during semantic-ID construction.
SAHL then uses exactly this hierarchy to define token-, subspace-, and item-level reconstruction targets along historical paths. An information-theoretic strategy~\cite{mackay2002information} prioritizes uncertain targets, while curriculum learning~\cite{bengio2009curriculum} transitions from random masking through entropy-guided reconstruction to uncorrupted next-item generation. In this way, SRBQ defines the units to reconstruct, and the learning strategies determine how those units supervise path understanding.
The joint design provides complementary benefits at both stages. By reconstructing the same token, subspace, and item units that SRBQ defines during quantization, SAHL grounds historical supervision in the representation structure rather than in arbitrary token blocks. Moreover, reconstructing informative units distributed across earlier interactions encourages the model to recover persistent preferences beyond local transitions, while curriculum optimization carries this ability into next-item generation.

Experiments on four Amazon Reviews 2014 categories~\cite{mcauley2015image} show that SAHL improves NDCG@10 over the architecture-matched RPG baseline by 18.6\%--30.0\%. SRBQ reduces mean/worst peak codeword loads by 28.8\%--32.4\%/37.3\%--41.3\%. Grouping controls show that reusing actual SRBQ groups outperforms same-size artificial groups, while history shuffling verifies that reconstruction uses user-specific context. Transfer to text tokens further separates the general history-learning mechanism from SRBQ-specific benefits. The contributions of this work are:

\begin{itemize}
    \item We formulate semantic-ID construction and history learning as a coupled design problem and propose SAHL, which reuses quantizer-defined ID structure to organize historical supervision.
    
    \item We introduce SRBQ, combining subspace decomposition, residual quantization, and occupancy-aware assignment to construct balanced token--subspace--item units for history reconstruction.
    \item We align masking with the SRBQ hierarchy, prioritize informative historical units through predictive entropy, and use curriculum learning to connect history reconstruction with future-path generation.
    \item Extensive experiments on four datasets demonstrate consistent gains and separately verify balanced reconstruction targets, the benefit of reusing quantizer-defined groups, and dependence on user history.
\end{itemize}

\section{Related Work}
\subsection{Generative Recommendation}
Sequential recommenders model ordered user behavior using Markov chains~\cite{rendle2010factorizing}, recurrent or convolutional networks~\cite{hidasi2016session,li2017neural,yue2024linear,tang2018personalized}, graph-based models~\cite{wu2019session,chang2021sequential}, and self-attention~\cite{kang2018self,sun2019bert4rec,hao2023feature}. Generative recommendation instead formulates recommendation as identifier generation~\cite{tay2022transformer,wang2022neural}. Language-model-based methods such as P5~\cite{geng2022p5} adopt text-to-text interfaces with prompting or efficient tuning~\cite{gao2023chatrec,jesse2023leveraging,bao2023tallrec}, while semantic-ID methods represent items as discrete tuples to enable exact catalog-level generation~\cite{rajput2023recommender}. Recent work advances identifier-based recommendation through representations, behavior, and decoding: GNPR-SID constructs semantic and collaborative identifiers for next-POI generation~\cite{wang2025gnpr}, EAGER jointly generates behavioral and semantic tokens~\cite{wang2024eager}, and RPG scales semantic-ID length through parallel decoding~\cite{hou2025generating}. IMBGen further combines residual and parallel quantization for multi-behavior generation~\cite{tan2025imbgen}, whereas LCKD-SR studies efficient knowledge transfer from LLMs to sequential recommenders~\cite{ao2026lckd}. Complementary directions include masked diffusion~\cite{mu2026mdgr}, cascaded ranking~\cite{lin2026care}, test-time reranking~\cite{wang2026isd}, and \mbox{collaborative-target alignment}~\cite{lin2026tca}.

Despite these advances, most methods focus on generation or inference, while supervision for modeling observed interaction histories remains comparatively underexplored. History self-supervision reconstructs masked items~\cite{sun2019bert4rec}, pretrains multiple sequence views~\cite{zhou2020s3rec}, or contrasts augmented histories~\cite{xie2022contrastive}, but is mainly developed for discriminative recommendation. MHL~\cite{wei-etal-2026-past} extends history reconstruction to causal semantic-ID generation. SAHL aligns history supervision with SRBQ's token-, subspace-, and item-level hierarchy.

\subsection{Semantic ID Construction and Vector Quantization}

Item identifiers define the output space and the semantic relationships exposed to the recommender. Atomic IDs contain no explicit content similarity, whereas modality encoders map item descriptions and attributes into dense representations~\cite{yuan2023where,hou2023learning}; Sentence-T5~\cite{Muresan2022SentenceT5}, for example, places semantically related items close in the embedding space.

Vector quantization converts these dense representations into discrete codes suitable for generation. Product quantization~\cite{jegou2011pq} independently encodes low-dimensional subvectors, while OPQ~\cite{ge2013optimized} learns a global rotation to reduce quantization distortion under a fixed code budget. Residual quantization instead successively encodes the information left unexplained by previous codewords, yielding a coarse-to-fine representation~\cite{lee2022rqvae}. RQ-VAE demonstrates that stacked residual codes support both reconstruction and autoregressive generation, and TIGER adopts this formulation for recommendation~\cite{rajput2023recommender}. Subsequent studies explore transferable quantized representations~\cite{hou2023learning}, compact indexing for large catalogs~\cite{petrov2024recjpq}, and learnable or task-aware code assignments~\cite{wang2024learnable,qu2024tokenrec}.

More recent methods further refine semantic-ID construction. DIGER~\cite{fu2026diger} exposes code assignments to recommendation gradients; ReSID~\cite{liang2026resid} jointly considers predictability, discrimination, and balance; DOS~\cite{yin2026dos} combines collaborative alignment with orthogonal residual quantization; CapsID~\cite{cheng2026capsid} introduces variable-length identifiers; and AsymRec~\cite{huang2026asymrec} constructs multi-faceted hierarchies. Tlow~\cite{li2026tlow} transforms Sentence-T5 embeddings before partitioned quantization and aligns token embeddings with codebook geometry. These methods increasingly emphasize code distribution, utilization, and hierarchy. SAHL complements this direction by further using the internal structure of semantic IDs to define aligned history-reconstruction and uncertainty units. 
In this work, SRBQ groups decorrelated components, refines each group residually, and constrains peak occupancy. SAHL uses the resulting \emph{token} $\subset$ \emph{subspace} $\subset$ \emph{item} hierarchy as aligned reconstruction units, coupling representation and objective.

\section{Preliminaries and Problem Formulation}
\label{sec:preliminaries}

\subsection{Problem Definition and Semantic ID Representation}
\label{sec:preliminary}

Let $\mathcal{U}$ and $\mathcal{I}$ be the user and item sets. Given user $u$'s chronological history $\boldsymbol{S}_u=(i_1,\ldots,i_T)$, the task is to generate $i_{T+1}$. Item $i$'s available metadata are concatenated into $a_i$ and encoded with Sentence-T5~\cite{Muresan2022SentenceT5}:
\begin{equation}
\mathbf{x}_i
=
\operatorname{SentenceT5}(a_i)
\in\mathbb{R}^{d_x},
\label{eq:item_text_embedding}
\end{equation}
where $d_x$ is the Sentence-T5 embedding dimensionality.

A quantizer maps $\mathbf{x}_i$ to $K$ tokens. Let $\mathcal{Q}^k=\{1,\ldots,M\}$ be codebook $k$'s local indices and $\mathcal{V}^k$ its disjoint emitted-token set. An offset $o_k$ maps local assignment $q_i^k$ to $z_i^k=o_k(q_i^k)$, giving
\begin{equation}
\phi(i)=(z_i^1,z_i^2,\ldots,z_i^K),\qquad z_i^k\in\mathcal{V}^k,
\end{equation}
where $q_i^k$ is the local codeword index and $z_i^k$ is its position-specific token in $\mathcal{V}=\bigcup_k\mathcal{V}^k$. The tokenized history is $\boldsymbol{Z}_u=(\phi(i_1),\ldots,\phi(i_T))$. Hereafter, \emph{token} means a semantic token; ordinary language tokens are \emph{text tokens}.

The $K$ token embeddings of $i_t$ are mean-pooled and encoded by a causal Transformer as $\mathbf{h}_t$. Position-specific head $f_k$ produces $\mathbf{h}_t^k$, and token $v\in\mathcal{V}^k$ with embedding $\mathbf{e}_v^k$ has probability
\begin{equation}
P_\theta^k(v\mid\mathbf{h}_t)
=
\frac{\exp\left(\operatorname{sim}(\mathbf{h}_t^k,\mathbf{e}_v^k)/\tau\right)}
{\sum_{v'\in\mathcal{V}^k}\exp\left(\operatorname{sim}(\mathbf{h}_t^k,\mathbf{e}_{v'}^k)/\tau\right)},
\label{eq:token_prob}
\end{equation}
where $\operatorname{sim}$ is cosine similarity, $\tau$ is the temperature, and $\theta$ denotes trainable parameters. These token probabilities form the common prediction interface used by next-item generation and history reconstruction.

\begin{figure*}[!t]
  \centering

  \includegraphics[width=\textwidth]{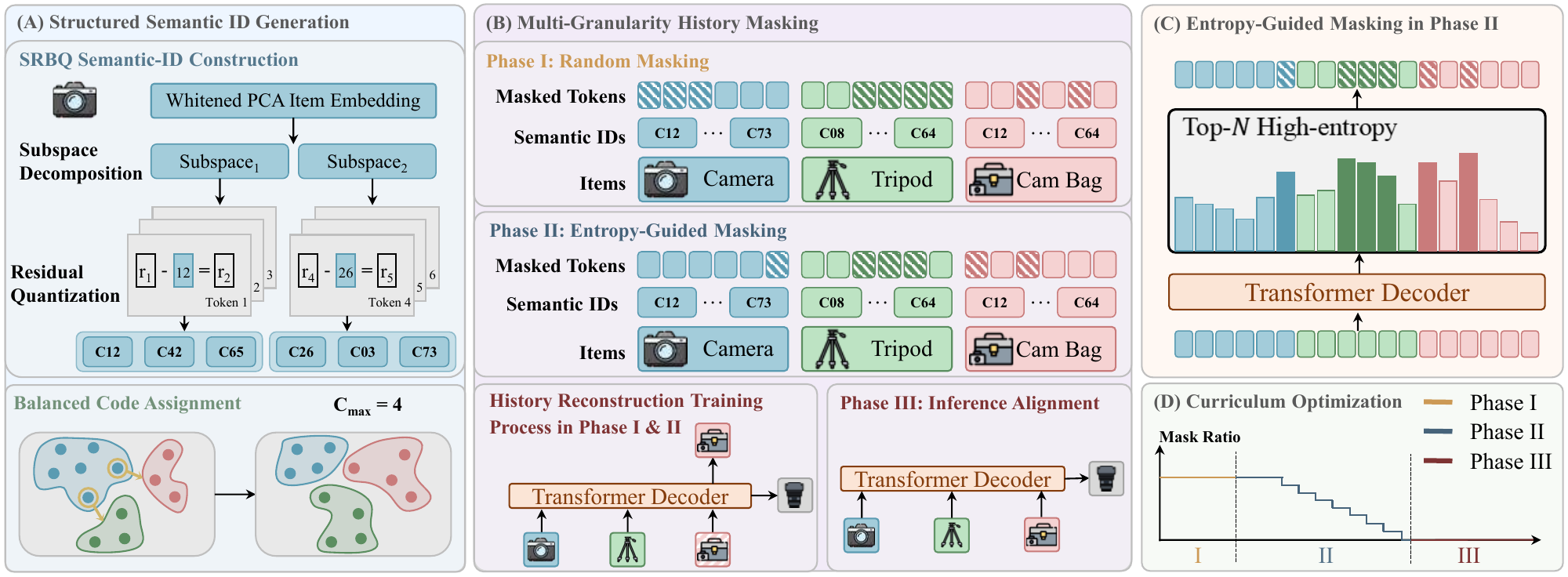}
  \caption{SAHL framework overview. (A) SRBQ constructs structured semantic IDs through subspace decomposition, intra-subspace residual quantization, and balanced code assignment. (B) Multi-granularity history masking reconstructs historical units at token, subspace, and item levels. (C) Entropy-guided masking selects uncertain historical targets. (D) Curriculum optimization progresses from random masking through entropy guidance to inference alignment.}
  \label{fig:framework}
\Description{SAHL framework connecting subspace-structured quantization, multi-granularity history reconstruction, entropy guidance, and curriculum optimization.}
\end{figure*}

\section{Structure-Aligned History Learning}
\label{sec:method}

\subsection{Framework Overview}

SAHL follows one design principle: structural units introduced during semantic-ID construction should remain available to the history-learning objective. Figure~\ref{fig:framework} organizes this principle into four components. (A) SRBQ constructs token, subspace, and item units; (B) multi-granularity masking reconstructs matching historical units; (C) entropy-guided masking prioritizes positions preceding uncertain transitions; and (D) curriculum optimization returns training gradually to uncorrupted prediction. Accordingly, we first describe semantic-ID generation, define the masking units and their learning objectives, and then introduce entropy-guided target selection and curriculum scheduling over those objectives.

\subsection{Semantic ID Generation}
\label{sec:srbq}

\subsubsection{OPQ-Based Semantic IDs}

For the baseline, whitened PCA reduces the Sentence-T5 embedding to $\bar{\mathbf{x}}_i\in\mathbb{R}^{d_{\mathrm{opq}}}$. OPQ~\cite{ge2013optimized} learns an orthogonal rotation $\mathbf{R}$ and partitions the result into $K$ subvectors,
\begin{equation}
\mathbf{R}\bar{\mathbf{x}}_i
=
\left[
\mathbf{x}_{i,\mathrm{rot}}^{(1)};
\ldots;
\mathbf{x}_{i,\mathrm{rot}}^{(K)}
\right],
\label{eq:opq_partition}
\end{equation}
and each subvector is independently assigned to its nearest codeword:
\begin{equation}
q_i^k
=
\arg\min_{m\in\{1,\ldots,M\}}
\left\|
\mathbf{x}_{i,\mathrm{rot}}^{(k)}-\mathbf{c}_{k,m}
\right\|_2^2,
\label{eq:opq_assignment}
\end{equation}
where $\mathbf{c}_{k,m}$ is codeword $m$ in the $k$-th OPQ codebook.
The identifier is $\phi_{\mathrm{OPQ}}(i)=[o_1(q_i^1),\ldots,o_K(q_i^K)]$. 

\subsubsection{Semantic ID Structure for Multi-Granularity Reconstruction}

To provide explicit reconstruction units, SAHL introduces \emph{Semantic Residual Balanced Quantization} (SRBQ). SRBQ decomposes decorrelated dimensions into $G$ semantic subspaces, applies $D$ residual-quantization stages per subspace, and constrains local codeword occupancy. It thereby organizes the $K=GD$ tokens by semantic group and residual level:
\begin{equation}
\text{token}\ \subset\ \text{subspace}\ \subset\ \text{item}.
\label{eq:srbq_hierarchy}
\end{equation}

This hierarchy defines SAHL's token-, subspace-, and item-level reconstruction units. In contrast, OPQ independently quantizes rotated subvectors for compact distortion-oriented representation, but its token positions form no coherent intermediate unit between a token and an item. Moreover, distortion minimization does not regulate local codeword occupancy, allowing some codes to remain sparsely trained while others become overloaded. SRBQ addresses these limitations through the three operations formalized below.

\subsubsection{Semantic Subspace Decomposition}

SRBQ first applies whitened principal component analysis to the item embeddings:
\begin{equation}
\widetilde{\mathbf{x}}_i
=
\operatorname{PCA}(\mathbf{x}_i)
\in\mathbb{R}^{d_{\mathrm{srbq}}},
\end{equation}
where $d_{\mathrm{srbq}}$ is the retained dimensionality. SRBQ then partitions the principal components into $G$ non-overlapping groups,
\begin{equation}
\widetilde{\mathbf{x}}_i
=
\left[
\widetilde{\mathbf{x}}_i^{(1)};
\widetilde{\mathbf{x}}_i^{(2)};
\ldots;
\widetilde{\mathbf{x}}_i^{(G)}
\right].
\label{eq:subspace_partition}
\end{equation}

Dimensions are divided as evenly as possible when $d_{\mathrm{srbq}}$ is not divisible by $G$. Whitening removes scale differences, while PCA decorrelates dimensions; grouping the resulting components therefore provides complementary latent views rather than arbitrary segments of the original embedding. SRBQ chooses $G$ such that $G\mid K$ and assigns $D=K/G$ consecutive positions to each group. With $k(g,\ell)=(g-1)D+\ell$, the local assignment is $q_i^{g,\ell}=q_i^{k(g,\ell)}$, the emitted token is $z_i^{g,\ell}=o_{k(g,\ell)}(q_i^{g,\ell})$, and centers are stored in $\mathcal{C}_{g,\ell}$. Thus, $G$ controls the semantic views and $D$ their residual depth.

\subsubsection{Intra-Subspace Residual Quantization}

Within each subspace, SRBQ uses $D$ successive quantizers to encode complementary residual information. For group $g$, the initial residual is
\begin{equation}
\mathbf{r}_{i,g}^{1}=\widetilde{\mathbf{x}}_i^{(g)}.
\end{equation}

At residual level $\ell\in\{1,\ldots,D\}$, K-means produces a codebook $\mathcal{C}_{g,\ell}=\{\mathbf{c}_{g,\ell,1},\ldots,\mathbf{c}_{g,\ell,M}\}$. Item $i$ is assigned to the nearest center,
\begin{equation}
q_i^{g,\ell}
=
\arg\min_{m\in\{1,\ldots,M\}}
\left\|
\mathbf{r}_{i,g}^{\ell}-\mathbf{c}_{g,\ell,m}
\right\|_2^2,
\label{eq:srbq_assignment}
\end{equation}
and the unexplained component is passed to the next level:
\begin{equation}
\mathbf{r}_{i,g}^{\ell+1}
=
\mathbf{r}_{i,g}^{\ell}
-
\mathbf{c}_{g,\ell,q_i^{g,\ell}}.
\label{eq:srbq_residual}
\end{equation}

Later positions encode information unexplained by earlier codewords, forming a coarse-to-fine description within each group. The corresponding ablation retains all positions and independent quantizers, isolating residual refinement.

\subsubsection{Balanced Code Assignment}

To avoid under-trained and overloaded codes, SRBQ caps cluster occupancy for catalog size $N=|\mathcal{I}|$:
\begin{equation}
C_{\max}
=
\max\left\{
\left\lceil\frac{N}{M}\right\rceil,
\left\lfloor\beta\frac{N}{M}\right\rfloor
\right\},
\label{eq:balance_capacity}
\end{equation}
where $\beta\geq1$ controls imbalance. The ceiling guarantees enough capacity when $N/M$ is noninteger, including at $\beta=1$; the second term supplies the ratio-controlled cap otherwise. Values near one approach uniform capacity, while larger values approach unconstrained assignment. After K-means, the farthest excess items in overflowing clusters move to their nearest under-capacity alternatives, followed by a few center updates. Only overflow assignments change, limiting peak load without enforcing exact uniformity.

\subsubsection{Structured Semantic Identifier}

The final semantic ID concatenates the residual codes group by group:
\begin{equation}
\phi_{\mathrm{SRBQ}}(i)
=
\left[
z_i^{1,1},\ldots,z_i^{1,D},
\ldots,
z_i^{G,1},\ldots,z_i^{G,D}
\right].
\label{eq:srbq_id}
\end{equation}

The $D$ tokens in each group jointly describe one latent subspace, and all $G$ groups represent the complete item. Each token therefore has two coordinates---its semantic group and residual level---making the subspace a meaningful reconstruction target rather than an arbitrary contiguous span. Recovering a group requires inferring its latent component from prior interactions and the groups still visible in the current item.

\subsection{Multi-Granularity History Masking}
\label{sec:structure_masking}

SAHL instantiates $\Omega$ at token, subspace, and item levels; subspace masking requires SRBQ groups. These levels remove a component, latent view, or complete interaction, respectively.

\subsubsection{Atomic Masking Granularities}

\paragraph{Token-Level Masking}
Masking $\Omega_{t,k}^{\mathrm{tok}}=\{(t,k)\}$ recovers a code from history and unmasked item tokens.

\paragraph{Subspace-Level Masking}
For subspace $g$, subspace-level masking jointly removes its $D$ residual tokens:
\begin{equation}
\Omega_{t,g}^{\mathrm{sub}}
=
\left\{
(t,k(g,\ell)):
1\leq \ell\leq D
\right\}.
\label{eq:subspace_mask}
\end{equation}

Reconstruction infers this PCA subspace's latent component from prior interactions and visible semantic groups.

\paragraph{Item-Level Masking}
Item masking selects $\Omega_t^{\mathrm{item}}=\{(t,k):1\leq k\leq K\}$, removing the complete item to emphasize inter-item dependencies. Under SRBQ it equals $\bigcup_{g=1}^{G}\Omega_{t,g}^{\mathrm{sub}}$.

\subsubsection{Composite Masking Strategies}

SAHL routes each valid historical item to an atomic branch:
\begin{equation}
\xi_t
\sim
\operatorname{Categorical}
\left(
\pi_{\mathrm{tok}},
\pi_{\mathrm{sub}},
\pi_{\mathrm{item}}
\right),
\label{eq:mixed_routing}
\end{equation}
where $\boldsymbol{\pi}$ is a probability vector. \textit{Mixed} disables subspace masking; \textit{Tri-Mixed} uses all branches. Routing precedes target selection, and $\rho$ selects units within each branch subject to rounding and a nonempty target.

\subsection{Learning Objectives}
\label{sec:learning_objectives}

Given the reconstruction set $\Omega$, SAHL jointly optimizes next-item generation and masked history reconstruction. Prediction is averaged over the non-ignored positions supplied by the tokenizer. Reconstruction losses are averaged within each active prediction head and then across heads, preventing heavily masked heads from dominating the update.

\paragraph{Next-Item Prediction}

Let $\mathcal{P}$ denote the non-ignored prediction positions supplied for optimization. For an update with nonempty $\Omega$, masking precedes the shared Transformer pass and prediction uses the corrupted state $\widetilde{\mathbf{h}}_t$; when $\Omega=\varnothing$, it uses the clean state $\mathbf{h}_t$. We denote this state by $\mathbf{h}_t^{\Omega}$. Each state predicts the $K$ tokens of $i_{t+1}$ in parallel, while inference uses the clean final state $\mathbf{h}_T$. The prediction loss is
\begin{equation}
\mathcal{L}_{\mathrm{pred}}
=
-\frac{1}{|\mathcal{P}|K}
\sum_{t\in\mathcal{P}}
\sum_{k=1}^{K}
\log
P_\theta^k
\left(
z_{i_{t+1}}^k
\mid
\mathbf{h}_t^{\Omega}
\right).
\label{eq:prediction_loss}
\end{equation}

\paragraph{Masked History Reconstruction}

Let $\widetilde{\boldsymbol{Z}}_u$ replace positions in $\Omega$ with position-aware mask tokens. With corrupted state $\widetilde{\mathbf{h}}_t$, $\Omega_k=\{t:(t,k)\in\Omega\}$, and active heads $\mathcal{K}_{\Omega}=\{k:|\Omega_k|>0\}$, the reconstruction loss is
\begin{equation}
\mathcal{L}_{\mathrm{rec}}
=
-\frac{1}{|\mathcal{K}_{\Omega}|}
\sum_{k\in\mathcal{K}_{\Omega}}
\frac{1}{|\Omega_k|}
\sum_{t\in\Omega_k}
\log
P_\theta^k
\left(
z_{i_t}^k
\mid
\widetilde{\mathbf{h}}_t
\right).
\label{eq:reconstruction_loss}
\end{equation}

The causal backbone reconstructs each target from the preceding path and the unmasked tokens at the current position. With nonempty reconstruction targets, the overall objective is
\begin{equation}
\mathcal{L}
=
\lambda_{\mathrm{pred}}\mathcal{L}_{\mathrm{pred}}
+
\lambda_{\mathrm{rec}}\mathcal{L}_{\mathrm{rec}},
\label{eq:overall_loss}
\end{equation}
where $\lambda_{\mathrm{pred}}$ and $\lambda_{\mathrm{rec}}$ weight the two tasks. When $\Omega=\varnothing$, $\mathcal{L}_{\mathrm{rec}}=0$ and the model optimizes next-item prediction alone. SRBQ is performed offline, entropy refresh requires one gradient-free pass, and inference uses neither operation.

\subsection{Granularity-Aware Entropy Guidance}
\label{sec:entropy_mask}

Predictive entropy locates historical positions followed by uncertain next-item transitions. The score at $t$ measures the difficulty of predicting $i_{t+1}$ from the prefix ending at $i_t$ and is assigned back to $t$ for reconstruction. It therefore measures transition difficulty rather than uncertainty about the observed item itself, encouraging the model to revisit context preceding an ambiguous transition.

\subsubsection{Online Uncertainty Estimation}

An entropy-guided update first performs a gradient-free, dropout-disabled pass over the uncorrupted history. Let $\widetilde{P}_\theta^k$ follow Eq.~\eqref{eq:token_prob} with the same temperature $\tau$ under this deterministic pass. Then
\begin{equation}
\mathcal{H}_{t}^{k}
=
-\sum_{v\in\mathcal{V}^k}
\widetilde{P}_\theta^k(v\mid\mathbf{h}_t)
\log
\widetilde{P}_\theta^k(v\mid\mathbf{h}_t),
\label{eq:token_entropy}
\end{equation}
measures uncertainty of token $k$ of $i_{t+1}$. The tilde marks deterministic evaluation; cached scores are refreshed periodically.

\subsubsection{Hierarchical Entropy Aggregation}

Token entropies are aggregated with SRBQ's hierarchy:
\begin{equation}
\begin{aligned}
\mathcal{H}_{t,g}^{\mathrm{sub}}&=\operatorname{Agg}(\{\mathcal{H}_{t}^{k(g,\ell)}\}_{\ell=1}^{D}),\\
\mathcal{H}_{t}^{\mathrm{item}}&=\operatorname{Agg}(\{\mathcal{H}_{t}^{k}\}_{k=1}^{K}).
\end{aligned}
\label{eq:hierarchical_entropy}
\end{equation}

SAHL defaults to mean aggregation and also evaluates max aggregation over matching reconstruction units.

\subsubsection{Entropy-Guided Target Selection}

Let $\mathcal{A}_\nu$ contain candidates at granularity $\nu\in\{\mathrm{tok},\mathrm{sub},\mathrm{item}\}$. Candidates are $(t,k)$, $(t,g)$, and $t$, respectively, with scores $\mathcal{H}_{\mathrm{tok}}(t,k)=\mathcal{H}_t^k$, $\mathcal{H}_{\mathrm{sub}}(t,g)=\mathcal{H}_{t,g}^{\mathrm{sub}}$, and $\mathcal{H}_{\mathrm{item}}(t)=\mathcal{H}_t^{\mathrm{item}}$. Let $\Gamma_\nu(c)$ expand candidate $c$ into reconstruction indices. Ratio $\rho$ selects
\begin{equation}
n_\nu
=
\min
\left(
|\mathcal{A}_\nu|,
\max
\left(
n_\nu^{\min},
\operatorname{round}
\left(
\rho|\mathcal{A}_\nu|
\right)
\right)
\right),
\end{equation}
where $n_\nu^{\min}$ prevents an empty target whenever $\mathcal{A}_\nu$ is nonempty. The highest-entropy candidates are selected and expanded into reconstruction positions as
\begin{equation}
\mathcal{S}_\nu
=
\underset{\substack{\mathcal{S}\subseteq\mathcal{A}_\nu\\|\mathcal{S}|=n_\nu}}{\arg\max}
\sum_{c\in\mathcal{S}}\mathcal{H}_\nu(c),
\label{eq:entropy_topk}
\end{equation}
and
\begin{equation}
\Omega_\nu
=
\bigcup_{c\in\mathcal{S}_\nu}\Gamma_\nu(c).
\end{equation}

The final target set is $\Omega=\bigcup_{\nu\in\{\mathrm{tok},\mathrm{sub},\mathrm{item}\}}\Omega_\nu$, with inactive branches contributing an empty set. Composite strategies first route items by Eq.~\eqref{eq:mixed_routing}; \textit{Entropy-Mixed} disables subspaces and \textit{Entropy-Tri-Mixed} retains all branches. Separate branch selection avoids ranking unlike units together.

\subsection{Curriculum Optimization}
\label{sec:curriculum}

Entropy is least reliable before the model learns a meaningful prediction space, while persistent corruption mismatches the uncorrupted histories seen at inference. The curriculum addresses both issues: it first learns stable sequence and reconstruction features without uncertainty, subsequently targets difficult transitions, and finally removes the auxiliary task so late training matches deployment.

\paragraph{Phase I: Granularity-Matched Random Warm-Up}
Training uses the requested strategy's random counterpart at the configured ratio $\rho_0=\rho$, establishing prediction and reconstruction before entropy selection.

\paragraph{Phase II: Entropy-Guided Reconstruction with Adaptive Decay}
After a minimum warm-up and a validation plateau, training switches to entropy guidance at the initial ratio. Once the entropy-guided stage reaches a plateau, it enters an adaptive-decay segment in which each subsequent evaluation reduces the ratio:
\begin{equation}
\rho_j
=
\max
\left(
0,
\rho_{j-1}-\delta\rho_0
\right),
\label{eq:linear_mask_decay}
\end{equation}
where $\delta$ is the decay fraction. Decreasing $\rho_j$ shifts optimization toward next-item generation.

\paragraph{Phase III: Inference Alignment}
At $\rho_j=0$, reconstruction is disabled and training matches uncorrupted inference.

\begin{table}[!t]
\centering
\caption{Dataset statistics; Avg. Length is the mean sequence length.}
\label{tab:dataset_stats}
\begin{tabular*}{\linewidth}{@{\extracolsep{\fill}}lcccc@{}}
\toprule
Dataset & \#Users & \#Items & \#Interactions & Avg. Length \\
\midrule
Beauty & 22,363 & 12,101 & 198,502 & 8.88 \\
Toys and Games & 19,412 & 11,924 & 167,597 & 8.63 \\
Sports and Outdoors & 35,598 & 18,357 & 296,337 & 8.32 \\
CDs and Vinyl & 75,258 & 64,443 & 1,097,592 & 14.58 \\
\bottomrule
\end{tabular*}
\end{table}

\section{Experimental Setup}
\label{sec:experimental_setup}

\subsection{Datasets}

We evaluate on four product categories from Amazon Reviews 2014~\cite{mcauley2015image}: \textit{Beauty}, \textit{Toys and Games}, \textit{Sports and Outdoors}, and \textit{CDs and Vinyl}. Following prior generative recommendation studies~\cite{rajput2023recommender,hou2025generating}, we apply core-5 filtering~\cite{he2016ups}, retaining only users and items with at least five interactions. For each item, we concatenate its available title, price, brand, feature, category, and description fields into a natural-language description for semantic representation learning~\cite{wang2024learnable}. Table~\ref{tab:dataset_stats} reports the statistics after preprocessing.

\subsection{Evaluation Metrics}

We evaluate recommendation performance using Recall@$k$ and NDCG@$k$ with $k\in\{5,10\}$. Recall@$k$ measures whether the ground-truth item is retrieved in the top-$k$ list, whereas NDCG@$k$ additionally rewards higher-ranked hits. Following prior work~\cite{kang2018self,sun2019bert4rec,rajput2023recommender,hou2025generating}, we adopt the standard leave-one-out protocol: for each chronologically ordered user sequence, the last interaction is used for testing, the penultimate interaction for validation, and the remaining interactions for training. NDCG@10 is the primary metric in sensitivity and cross-configuration analyses.

\subsection{Baselines}

We compare SAHL with representative baselines in two categories. \textbf{Item-ID methods} operate directly on learned item embeddings: Caser~\cite{tang2018personalized}, GRU4Rec~\cite{hidasi2016session}, HGN~\cite{ma2019hierarchical}, SASRec~\cite{kang2018self}, BERT4Rec~\cite{sun2019bert4rec}, FDSA~\cite{hao2023feature}, and S$^3$-Rec~\cite{zhou2020s3rec}. \textbf{Semantic-ID and generative methods} represent or generate items through discrete identifiers: RecJPQ~\cite{petrov2024recjpq}, VQ-Rec~\cite{hou2023learning}, TIGER~\cite{rajput2023recommender}, HSTU~\cite{zhai2024actions}, RPG~\cite{hou2025generating}, Tlow~\cite{li2026tlow}, and MHL~\cite{wei-etal-2026-past}.

Among the closest comparisons, RPG provides the parallel semantic-ID generation backbone, MHL augments this backbone with masked-history learning over OPQ identifiers, and Tlow is a recent tokenizer-oriented generative recommender. Tlow results follow its published matched protocol. For controlled comparisons, the reproduced RPG baseline, MHL, and SAHL use the same GPT-2~\cite{radford2019language} backbone and Sentence-T5-base item representations~\cite{Muresan2022SentenceT5}.

\begin{table*}[!t]
\centering
\caption{Comparison with existing recommenders. Best results are \textbf{bold}; second-best results are \underline{underlined}. $^*$ denotes Sentence-T5-base RPG reproduction with Direct Inference.}
\label{tab:all_models_transposed}
\small
\begin{tabular*}{\linewidth}{@{\extracolsep{\fill}}lrrrrrrrr@{}}
\toprule
\textbf{Model} & \multicolumn{4}{c}{\textbf{Beauty}} & \multicolumn{4}{c}{\textbf{Toys and Games}} \\
 & R@5 & N@5 & R@10 & N@10 & R@5 & N@5 & R@10 & N@10 \\
\midrule

\multicolumn{9}{c}{\textbf{Item ID-based}} \\
\midrule
Caser & .0205 & .0131 & .0347 & .0176 & .0166 & .0107 & .0270 & .0141 \\
GRU4Rec & .0164 & .0099 & .0283 & .0137 & .0097 & .0059 & .0176 & .0084 \\
HGN & .0325 & .0206 & .0512 & .0266 & .0321 & .0221 & .0497 & .0277 \\
BERT4Rec & .0203 & .0124 & .0347 & .0170 & .0116 & .0071 & .0203 & .0099 \\
SASRec & .0387 & .0249 & .0605 & .0318 & .0463 & .0306 & .0675 & .0374 \\
FDSA & .0267 & .0163 & .0407 & .0208 & .0228 & .0140 & .0381 & .0189 \\
S$^3$-Rec & .0387 & .0244 & .0647 & .0327 & .0443 & .0294 & .0700 & .0376 \\
\midrule
\multicolumn{9}{c}{\textbf{Semantic ID-based}} \\
\midrule
RecJPQ & .0311 & .0167 & .0482 & .0222 & .0331 & .0182 & .0484 & .0231 \\
VQ-Rec & .0457 & .0317 & .0664 & .0383 & .0497 & .0346 & .0737 & .0423 \\
TIGER & .0454 & .0321 & .0648 & .0384 & .0521 & .0371 & .0712 & .0432 \\
HSTU & .0469 & .0314 & .0704 & .0389 & .0433 & .0281 & .0669 & .0357 \\
RPG$^*$ & .0500 & .0358 & .0745 & .0436 & .0550 & .0386 & .0778 & .0460 \\
Tlow & .0545 & .0377 & .0786 & .0454 & .0590 & .0395 & .0864 & .0482 \\
MHL & \underline{.0574} & \underline{.0424} & \underline{.0795} & \underline{.0495} & \underline{.0672} & \underline{.0489} & \underline{.0903} & \underline{.0564} \\
\midrule
\textbf{SAHL} & \textbf{.0605} & \textbf{.0439} & \textbf{.0848} & \textbf{.0517} & \textbf{.0703} & \textbf{.0514} & \textbf{.0964} & \textbf{.0598} \\
\bottomrule
\end{tabular*}
\par\medskip
\begin{tabular*}{\linewidth}{@{\extracolsep{\fill}}lrrrrrrrr@{}}
\toprule
\textbf{Model} & \multicolumn{4}{c}{\textbf{Sports and Outdoors}} & \multicolumn{4}{c}{\textbf{CDs and Vinyl}} \\
 & R@5 & N@5 & R@10 & N@10 & R@5 & N@5 & R@10 & N@10 \\
\midrule

\multicolumn{9}{c}{\textbf{Item ID-based}} \\
\midrule
Caser & .0116 & .0072 & .0194 & .0097 & .0116 & .0073 & .0205 & .0101 \\
GRU4Rec & .0129 & .0086 & .0204 & .0110 & .0195 & .0120 & .0353 & .0171 \\
HGN & .0189 & .0120 & .0313 & .0159 & .0259 & .0153 & .0467 & .0220 \\
BERT4Rec & .0115 & .0075 & .0191 & .0099 & .0326 & .0201 & .0547 & .0271 \\
SASRec & .0233 & .0154 & .0350 & .0192 & .0351 & .0177 & .0619 & .0263 \\
FDSA & .0182 & .0122 & .0288 & .0156 & .0226 & .0137 & .0378 & .0186 \\
S$^3$-Rec & .0251 & .0161 & .0385 & .0204 & .0213 & .0130 & .0375 & .0182 \\
\midrule
\multicolumn{9}{c}{\textbf{Semantic ID-based}} \\
\midrule
RecJPQ & .0141 & .0076 & .0220 & .0102 & .0075 & .0046 & .0138 & .0066 \\
VQ-Rec & .0208 & .0144 & .0300 & .0173 & .0352 & .0238 & .0520 & .0292 \\
TIGER & .0264 & .0181 & .0400 & .0225 & .0492 & .0329 & .0748 & .0411 \\
HSTU & .0258 & .0165 & .0414 & .0215 & .0417 & .0275 & .0638 & .0346 \\
RPG$^*$ & .0284 & .0197 & .0436 & .0246 & .0463 & .0317 & .0668 & .0383 \\
Tlow & .0307 & .0207 & .0477 & .0261 & \underline{.0541} & .0362 & \underline{.0801} & .0446 \\
MHL & \underline{.0359} & \underline{.0249} & \underline{.0511} & \underline{.0298} & .0539 & \underline{.0369} & .0787 & \underline{.0449} \\
\midrule
\textbf{SAHL} & \textbf{.0374} & \textbf{.0260} & \textbf{.0547} & \textbf{.0315} & \textbf{.0555} & \textbf{.0376} & \textbf{.0820} & \textbf{.0461} \\
\bottomrule
\end{tabular*}
\end{table*}

\subsection{Implementation Details}

Our implementations and controlled experiments use a 2-layer, 4-head GPT-2 with hidden/feed-forward dimensions 448/1024; embedding/attention dropout is 0.3 and residual dropout is 0. AdamW uses zero weight decay, cosine scheduling, 10,000 warm-up steps, gradient clipping at 1.0, and batch size 64. In Table~\ref{tab:dataset_stats} order, learning rates are 0.01/0.003/0.003/0.001. Sentence-T5-base yields 768-dimensional embeddings; OPQ uses whitened PCA to 128 dimensions. ID lengths are 16/32/32/64 with $M=256$. OPQ uses Faiss \texttt{OPQ$K$,IVF1,PQ$K$x8} defaults; SRBQ uses 10 K-means initializations, at most 300 iterations, and 3 reassignment passes. Selected $(d_{\mathrm{srbq}},G,\beta)$ values are $(384,4,1.5)$, $(448,8,1.5)$, $(256,8,1.5)$, and $(384,16,1.2)$. Beam widths, graph neighborhoods, and propagation depths are 100/200/100/50, 100/100/200/500, and 3/5/2/4; temperature is 0.03. Controls use these settings.

Each item uses the mean log-probability of its $K$ tokens. Items sharing an ID remain separate candidates with equal scores, and ties are deterministically broken by fixed catalog indices. OPQ and SRBQ use the same graph-constrained top-$k$ decoder and no popularity or test information.

\section{Experimental Results}
\label{sec:experimental_results}

\subsection{Overall Performance Comparison}

Table~\ref{tab:all_models_transposed} compares SAHL with existing methods. RPG$^*$, MHL, and SAHL use the common setup in Section~\ref{sec:experimental_setup}; $^*$ denotes our Sentence-T5-base RPG reproduction with Direct Inference. Two-sided paired $t$-tests on per-user NDCG@10 yield $p<1\times10^{-10}$ against RPG and $p<7.5\times10^{-2}$ against MHL. Relative changes are calculated from the values displayed in the tables. From the results, we find:
SAHL leads every metric, improving NDCG@10 over RPG by 18.6\%--30.0\%, Tlow by 3.4\%--24.1\%, and MHL by 2.7\%--6.0\%. Relative to RPG, these gains reflect the combined effect of redesigning the representation side with SRBQ and the learning side with history reconstruction. The comparison with MHL further isolates the benefit of connecting them by reconstructing the groups defined during SRBQ quantization. Subsequent controls examine ID quality, grouping correspondence, and history.

\begin{table*}[!t]
\centering
\caption{Curriculum ablation under OPQ with $K=16$, token-level masking, $\rho=0.15$, and $\lambda_{\mathrm{rec}}=1.0$. Best and second-best results are \textbf{bold} and \underline{underlined}.}
\label{tab:curriculum_ablation}
\small
\begin{tabular*}{\linewidth}{@{\extracolsep{\fill}}lrrrrrrrr@{}}
\toprule
\textbf{Curriculum Strategy} & \multicolumn{4}{c}{\textbf{Beauty}} & \multicolumn{4}{c}{\textbf{Toys and Games}} \\
 & R@5 & N@5 & R@10 & N@10 & R@5 & N@5 & R@10 & N@10 \\
\midrule

Direct Inference & .0500 & .0358 & .0745 & .0436 & .0550 & .0386 & .0778 & .0460 \\
\midrule
R & \underline{.0554} & .0379 & \textbf{.0797} & .0457 & \textbf{.0647} & .0402 & \textbf{.0914} & .0488 \\
E & .0491 & .0338 & .0715 & .0410 & .0232 & .0151 & .0359 & .0191 \\
R$\to$Inf & \underline{.0554} & \underline{.0404} & .0791 & \underline{.0480} & .0623 & \underline{.0452} & .0876 & \textbf{.0533} \\
E$\to$Inf & .0518 & .0374 & .0736 & .0445 & .0471 & .0322 & .0702 & .0396 \\
R$\to$E$\to$Inf & \textbf{.0574} & \textbf{.0424} & \underline{.0795} & \textbf{.0495} & \underline{.0644} & \textbf{.0456} & \underline{.0883} & \textbf{.0533} \\
\bottomrule
\end{tabular*}
\par\medskip
\begin{tabular*}{\linewidth}{@{\extracolsep{\fill}}lrrrrrrrr@{}}
\toprule
\textbf{Curriculum Strategy} & \multicolumn{4}{c}{\textbf{Sports and Outdoors}} & \multicolumn{4}{c}{\textbf{CDs and Vinyl}} \\
 & R@5 & N@5 & R@10 & N@10 & R@5 & N@5 & R@10 & N@10 \\
\midrule

Direct Inference & .0284 & .0197 & .0436 & .0246 & .0463 & .0317 & .0668 & .0383 \\
\midrule
R & .0321 & .0179 & \underline{.0482} & .0231 & .0193 & .0104 & .0296 & .0137 \\
E & .0110 & .0070 & .0171 & .0090 & .0179 & .0115 & .0279 & .0147 \\
R$\to$Inf & \textbf{.0337} & \underline{.0233} & \textbf{.0486} & \underline{.0281} & \textbf{.0475} & \underline{.0329} & \textbf{.0675} & \underline{.0394} \\
E$\to$Inf & .0264 & .0180 & .0402 & .0224 & .0438 & .0312 & .0609 & .0367 \\
R$\to$E$\to$Inf & \underline{.0334} & \textbf{.0239} & .0474 & \textbf{.0284} & \underline{.0474} & \textbf{.0332} & \underline{.0674} & \textbf{.0396} \\
\bottomrule
\end{tabular*}
\end{table*}

\subsection{MHL Background under OPQ}
\label{sec:mhl_ablation}

\begin{figure}[!t]
\centering
\includegraphics[width=\linewidth]{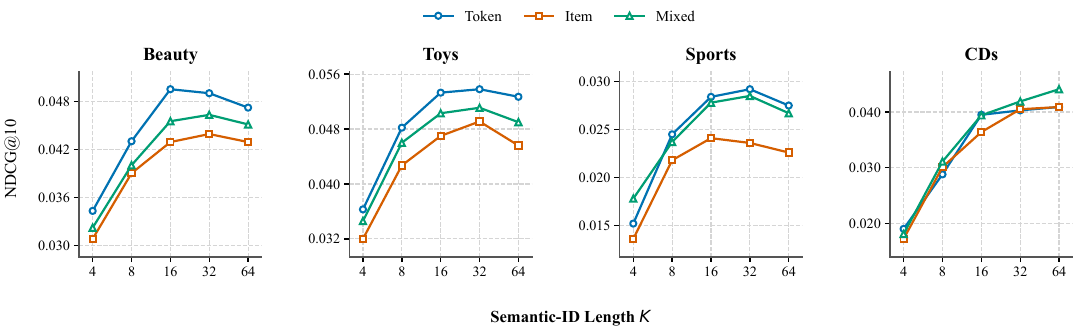}
\caption{NDCG@10 sensitivity to OPQ semantic-ID length $K$ under three masking strategies and R$\to$E$\to$Inf. The mask ratio is 0.15 for the first three datasets and 0.20 for CDs and Vinyl.}
\label{fig:ablation_codebook_size}
\Description{NDCG at 10 versus OPQ semantic-ID length for three masking strategies on four datasets.}
\end{figure}

We first study OPQ-based masked-history learning independently of SRBQ. Figures~\ref{fig:ablation_codebook_size}--\ref{fig:ablation_recon_loss} establish ID length, masking ratio, and reconstruction weight before coupling reconstruction to SRBQ-defined groups.

\subsubsection{Curriculum Strategy}

Table~\ref{tab:curriculum_ablation} compares six curricula with OPQ, $K=16$, token masking, $\rho=0.15$, and $\lambda_{\mathrm{rec}}=1.0$. Direct Inference is RPG-style training; R and E retain one reconstruction policy; R$\to$Inf and E$\to$Inf add inference alignment; R$\to$E$\to$Inf uses all three phases.

Reconstruction throughout training is unreliable: E trails Direct Inference everywhere, and R notably degrades performance on CDs and Vinyl. Entropy is weak from initialization because early predictive distributions do not yet provide reliable uncertainty. Schedules ending with Inf consistently recover performance; R$\to$E$\to$Inf gives the best NDCG@10 on Beauty, Sports and Outdoors, and CDs and Vinyl and ties on Toys and Games. R$\to$Inf provides the highest Recall on Sports and Outdoors and CDs and Vinyl. Overall, the results support random warm-up before targeted masking and a final inference-alignment phase.

\subsubsection{Semantic-ID Length}

Figure~\ref{fig:ablation_codebook_size} varies OPQ ID length $K\in\{4,8,16,32,64\}$ under three masks and the full curriculum. The mask ratio is 0.15 for the first three datasets and 0.20 for CDs and Vinyl. Increasing $K$ from 4 to 16 or 32 substantially improves the first three datasets, showing that short IDs lack semantic resolution; performance then saturates as longer targets fragment the generation space. The larger CDs and Vinyl catalog instead continues to benefit through $K=64$. We subsequently use $K=16/32/32/64$. Detailed results are given in Appendix~\ref{app:opq_details}, Table~\ref{tab:app_opq_length}.

\begin{figure}[!t]
\centering
\includegraphics[width=\linewidth]{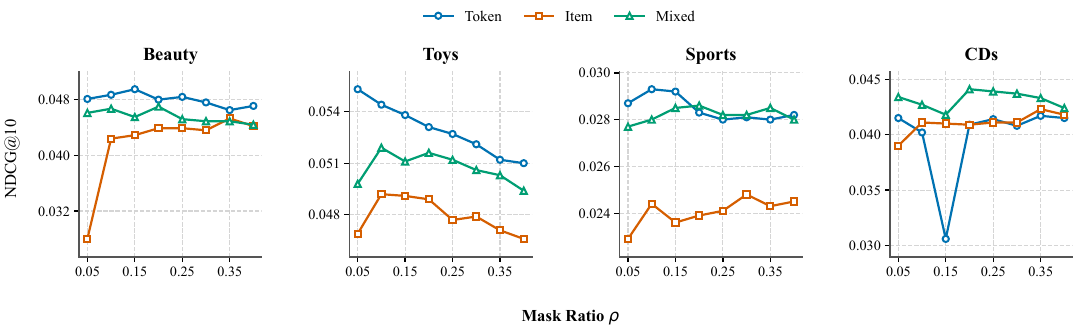}
\caption{NDCG@10 sensitivity to masking ratio $\rho$ under OPQ and R$\to$E$\to$Inf, using $K=16/32/32/64$ in Table~\ref{tab:dataset_stats} order.}
\label{fig:ablation_mask_ratio}
\Description{NDCG at 10 versus history masking ratio for three masking strategies on four datasets.}
\end{figure}

\subsubsection{Masking Ratio}

Figure~\ref{fig:ablation_mask_ratio} varies $\rho$ at each selected OPQ length. Token masking peaks at 0.15/0.05/0.10 on the first three datasets because it removes uncertain components while preserving most of the interaction path. Item masking is usually weaker because it removes all $K$ tokens. Mixed masking is most stable on CDs and Vinyl and peaks at 0.20 (.0441), suggesting that its larger catalog benefits from alternating fine- and coarse-grained reconstruction. Broad optima indicate limited sensitivity around these choices. Detailed results are given in Appendix~\ref{app:opq_details}, Table~\ref{tab:app_opq_ratio}.

\begin{figure}[!t]
\centering
\includegraphics[width=\linewidth]{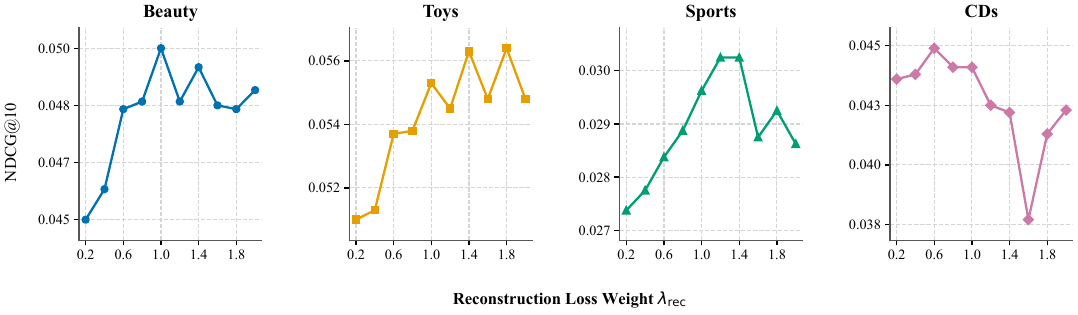}
\caption{NDCG@10 sensitivity to reconstruction weight $\lambda_{\mathrm{rec}}$ under the selected OPQ configuration of each dataset. The prediction weight is fixed at 1.0.}
\label{fig:ablation_recon_loss}
\Description{NDCG at 10 versus reconstruction-loss weight under selected OPQ settings on four datasets.}
\end{figure}

\subsubsection{Reconstruction Loss Weight}

Figure~\ref{fig:ablation_recon_loss} varies $\lambda_{\mathrm{rec}}$ with prediction weight 1.0 and the selected OPQ masks. The four datasets favor 1.0, 1.8, 1.2--1.4, and 0.6. Small weights underuse reconstruction, whereas large values hurt discrimination. Detailed results are given in Appendix~\ref{app:opq_details}, Table~\ref{tab:app_opq_loss}.

\begin{figure*}[!t]
\centering
\includegraphics[width=\textwidth]{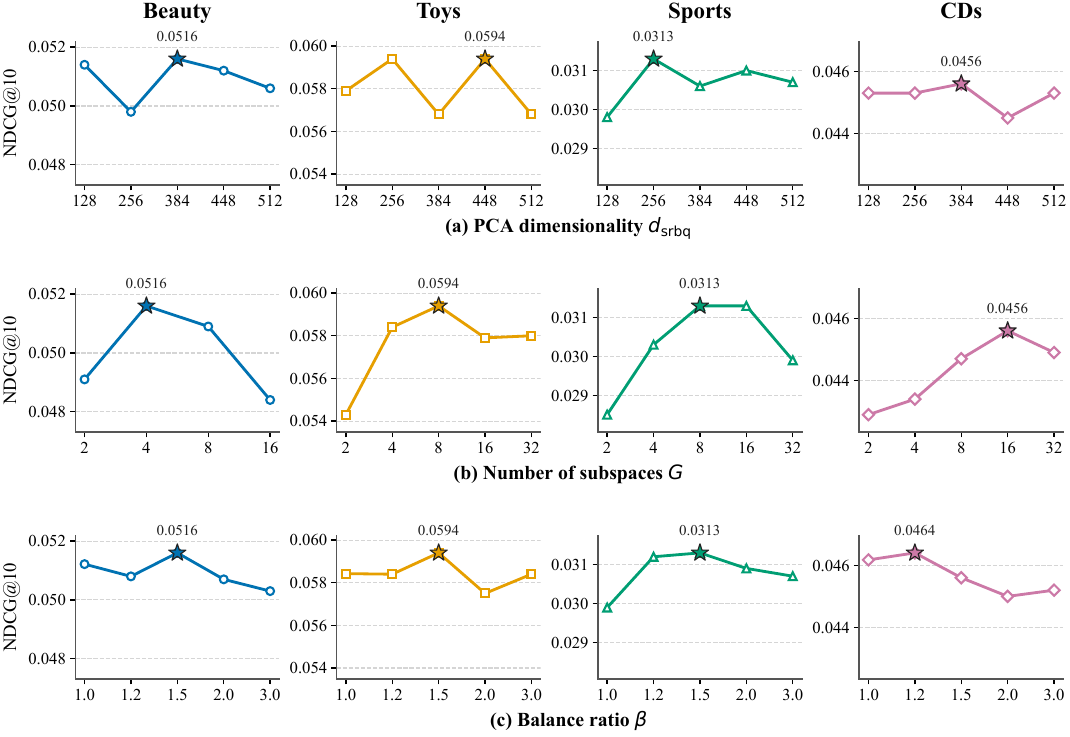}
\caption{SRBQ NDCG@10 sensitivity to (a) $d_{\mathrm{srbq}}$, (b) $G$, and (c) $\beta$. Stars mark selected values.}
\label{fig:srbq_parameter_search}
\Description{Sensitivity of NDCG at 10 to SRBQ embedding dimension, subspace count, and balance weight; stars indicate selected configurations.}
\end{figure*}

\subsection{SRBQ Configuration and Codebook Quality}

\subsubsection{SRBQ Parameter Search}

Figure~\ref{fig:srbq_parameter_search} sequentially tunes $d_{\mathrm{srbq}}\in\{128,256,384,448,512\}$, $G\in\{2,4,8,16,32\}$ subject to $G\le K$ and $G\mid K$, and $\beta\in\{1.0,1.2,1.5,2.0,3.0\}$. The selected $(d_{\mathrm{srbq}},G,\beta)$ values are $(384,4,1.5)$, $(448,8,1.5)$, $(256,8,1.5)$, and $(384,16,1.2)$. Sports and Outdoors ties at $G=8/16$ on NDCG@10, so Recall@10 selects 8. Detailed results are reported in Appendix~\ref{app:srbq_details}, Tables~\ref{tab:app_srbq_dimension}--\ref{tab:app_srbq_balance}.

From the experiment results, we find three trends emerge: Performance is nonmonotonic in retained dimension: too few components discard useful variation, whereas extending to 512 generally introduces weak dimensions that complicate quantization. Every dataset also favors an interior $G$; too few groups merge heterogeneous factors, while too many shorten each residual chain because $D=K/G$ and fragment its representation. Finally, the first three datasets favor $\beta=1.5$, while CDs and Vinyl favors 1.2. Peak-load control therefore helps, but overly rigid reassignment can sacrifice fidelity. As tuning is sequential, the curves show conditional rather than exhaustive joint sensitivity.

\subsubsection{Cross-Dataset Codebook Assignment Diagnostics}

We test whether Fig.~\ref{fig:codebook_profile}'s pattern generalizes without sacrificing ID distinctiveness. Let $n_{s,m}$ count the $N$ items using code $m$ at position $s$; ideal load is $\bar n=N/M$.

We report mean and worst peak-to-ideal load:
\begin{equation}
\label{eq:peak_load_ratio}
\begin{aligned}
r_s &= \frac{\max_{1\leq m\leq M} n_{s,m}}{\bar n},
&\qquad
\bar r &= \frac{1}{K}\sum_{s=1}^{K}r_s,\\
r_{\max} &= \max_{1\leq s\leq K} r_s.
\end{aligned}
\end{equation}

Normalized occupancy entropy is
\begin{equation}
\label{eq:occupancy_entropy}
\begin{aligned}
H_s &=-\frac{1}{\log M}\sum_{m=1}^{M}p_{s,m}\log p_{s,m},\\
\bar H &=\frac{1}{K}\sum_{s=1}^{K}H_s,
&p_{s,m}&=\frac{n_{s,m}}{N}.
\end{aligned}
\end{equation}

Full-ID redundancy and pairwise collision are
\begin{equation}
\label{eq:full_id_collision}
R_{\mathrm{dup}}=1-\frac{U}{N},
\qquad
P_{\mathrm{coll}}
=\frac{\sum_j c_j(c_j-1)}{N(N-1)},
\end{equation}
where $U$ is the number of unique IDs and $c_j$ their multiplicities. We also report active-code ratio $A=(KM)^{-1}\sum_s|\{m:n_{s,m}>0\}|$ and unique-ID ratio $U/N$.

\begin{table}[!t]
\centering
\caption{Peak-load diagnostics for OPQ and selected SRBQ configurations (lower is better).}
\label{tab:codebook_quality}
\begin{tabular*}{\linewidth}{@{\extracolsep{\fill}}lcccc@{}}
\toprule
\multirow{2}{*}{\textbf{Dataset}}
& \multicolumn{2}{c}{\textbf{Mean Peak Load $\bar r$ $\downarrow$}}
& \multicolumn{2}{c}{\textbf{Worst Peak Load $r_{\max}$ $\downarrow$}} \\
\cmidrule(lr){2-3}\cmidrule(lr){4-5}
& OPQ & SRBQ & OPQ & SRBQ \\
\midrule
Beauty & 2.19 & \textbf{1.48} & 2.52 & \textbf{1.48} \\
Toys and Games      & 2.08 & \textbf{1.48} & 2.36 & \textbf{1.48} \\
Sports and Outdoors & 2.12 & \textbf{1.49} & 2.47 & \textbf{1.49} \\
CDs and Vinyl       & 1.70 & \textbf{1.20} & 2.01 & \textbf{1.20} \\
\bottomrule
\end{tabular*}
\end{table}

Table~\ref{tab:codebook_quality} shows 28.8\%--32.4\% lower mean peaks and 37.3\%--41.3\% lower worst peaks. All codes remain active and unique-ID ratios are 99.68\%--99.97\%. On CDs and Vinyl, $(\bar r,r_{\max})$ falls from $(1.70,2.01)$ to $(1.20,1.20)$ and entropy rises from .990 to .999; redundancy (.296\%/.299\%) and collision probability ($4.8507/4.8521\times10^{-6}$) barely change.

The statistics capture complementary failures: $\bar r$ measures identifier-wide concentration and $r_{\max}$ positional bottlenecks. Unchanged coverage and collision behavior show genuine redistribution, not discarded codes or merged IDs, and yield more even support across reconstruction targets.

\begin{table*}[!t]
\centering
\caption{SRBQ component ablation using each dataset's selected quantization and history-learning configuration. Only the indicated component is removed; the remaining settings are fixed. Best results \textbf{bold}, second-best \underline{underlined}.}
\label{tab:srbq_component}
\small
\begin{tabular*}{\linewidth}{@{\extracolsep{\fill}}lrrrrrrrr@{}}
\toprule
\textbf{Variant} & \multicolumn{4}{c}{\textbf{Beauty}} & \multicolumn{4}{c}{\textbf{Toys and Games}} \\
 & R@5 & N@5 & R@10 & N@10 & R@5 & N@5 & R@10 & N@10 \\
\midrule

Full SRBQ & \textbf{.0605} & \textbf{.0439} & \textbf{.0848} & \textbf{.0517} & \textbf{.0703} & \textbf{.0514} & \textbf{.0964} & \textbf{.0598} \\
$-$ Balance & \underline{.0569} & \underline{.0416} & \underline{.0787} & \underline{.0487} & \underline{.0677} & \underline{.0478} & \underline{.0943} & \underline{.0564} \\
$-$ Residual & .0484 & .0350 & .0680 & .0413 & .0639 & .0449 & .0917 & .0538 \\
$-$ Subspace & .0558 & .0407 & .0771 & .0476 & .0631 & .0473 & .0838 & .0539 \\
\bottomrule
\end{tabular*}
\par\medskip
\begin{tabular*}{\linewidth}{@{\extracolsep{\fill}}lrrrrrrrr@{}}
\toprule
\textbf{Variant} & \multicolumn{4}{c}{\textbf{Sports and Outdoors}} & \multicolumn{4}{c}{\textbf{CDs and Vinyl}} \\
 & R@5 & N@5 & R@10 & N@10 & R@5 & N@5 & R@10 & N@10 \\
\midrule

Full SRBQ & \textbf{.0374} & \textbf{.0260} & \textbf{.0547} & \textbf{.0315} & \textbf{.0555} & \textbf{.0376} & \textbf{.0820} & \textbf{.0461} \\
$-$ Balance & \underline{.0365} & \underline{.0256} & \underline{.0515} & \underline{.0304} & \underline{.0551} & \underline{.0374} & \underline{.0798} & \underline{.0454} \\
$-$ Residual & .0282 & .0207 & .0410 & .0248 & .0533 & .0367 & .0773 & .0444 \\
$-$ Subspace & .0306 & .0222 & .0432 & .0263 & .0485 & .0330 & .0713 & .0403 \\
\bottomrule
\end{tabular*}
\end{table*}

\subsection{SRBQ Component Ablation}
\label{sec:srbq_component}

Table~\ref{tab:srbq_component} removes balanced assignment, residual updates (while retaining independent quantizers), or subspace decomposition (one PCA group). For each dataset, the full model's selected SRBQ configuration, masking strategy, masking ratio, semantic-ID length, and reconstruction-loss weight are retained except for the component being removed. Every removal lowers NDCG@10, supporting the contribution of all three components.

Removing residual coding reduces NDCG@10 by 3.7\%--21.3\% and causes the largest decline on Beauty, Toys and Games, and Sports and Outdoors. Every token position retains an independently trained quantizer, so the loss reflects missing progressive refinement rather than fewer tokens or parameters. On CDs and Vinyl, removing subspaces causes the largest decline, from .0461 to .0403 (12.6\%). Subspace removal reduces NDCG@10 by 7.9\%--16.5\% across datasets, showing that one residual chain cannot replace distinct semantic groups. Balance contributes a smaller but consistent 1.5\%--5.8\%, matching the load diagnostics. Together, the components define, refine, and balance reconstruction units.

\begin{table*}[!t]
\centering
\caption{Controlled quantizer and history-learning comparison under the common experimental configuration. The SRBQ+MHL row uses the complete dataset-specific settings and constitutes SAHL. Best results \textbf{bold}, second-best \underline{underlined}.}
\label{tab:quantizer_history_2x2}
\small
\begin{tabular*}{\linewidth}{@{\extracolsep{\fill}}llrrrrrrrr@{}}
\toprule
\textbf{Quantizer} & \textbf{Objective} & \multicolumn{4}{c}{\textbf{Beauty}} & \multicolumn{4}{c}{\textbf{Toys and Games}} \\
 &  & R@5 & N@5 & R@10 & N@10 & R@5 & N@5 & R@10 & N@10 \\
\midrule

OPQ & RPG & .0500 & .0358 & .0745 & .0436 & .0550 & .0386 & .0778 & .0460 \\
OPQ & MHL & \underline{.0574} & \underline{.0424} & .0795 & \underline{.0495} & \underline{.0672} & \underline{.0489} & \underline{.0903} & \underline{.0564} \\
SRBQ & RPG & .0536 & .0367 & \underline{.0796} & .0451 & .0555 & .0377 & .0825 & .0464 \\
SRBQ & MHL & \textbf{.0605} & \textbf{.0439} & \textbf{.0848} & \textbf{.0517} & \textbf{.0703} & \textbf{.0514} & \textbf{.0964} & \textbf{.0598} \\
\bottomrule
\end{tabular*}
\par\medskip
\begin{tabular*}{\linewidth}{@{\extracolsep{\fill}}llrrrrrrrr@{}}
\toprule
\textbf{Quantizer} & \textbf{Objective} & \multicolumn{4}{c}{\textbf{Sports and Outdoors}} & \multicolumn{4}{c}{\textbf{CDs and Vinyl}} \\
 &  & R@5 & N@5 & R@10 & N@10 & R@5 & N@5 & R@10 & N@10 \\
\midrule

OPQ & RPG & .0284 & .0197 & .0436 & .0246 & .0463 & .0317 & .0668 & .0383 \\
OPQ & MHL & \underline{.0359} & \underline{.0249} & \underline{.0511} & \underline{.0298} & \underline{.0539} & \underline{.0369} & \underline{.0787} & \underline{.0449} \\
SRBQ & RPG & .0296 & .0204 & .0455 & .0255 & .0472 & .0318 & .0696 & .0390 \\
SRBQ & MHL & \textbf{.0374} & \textbf{.0260} & \textbf{.0547} & \textbf{.0315} & \textbf{.0555} & \textbf{.0376} & \textbf{.0820} & \textbf{.0461} \\
\bottomrule
\end{tabular*}
\end{table*}

\subsection{Complementarity of SRBQ and Masked History Learning}

Table~\ref{tab:quantizer_history_2x2} controls quantizer and objective under the common setup. OPQ+RPG matches Table~\ref{tab:all_models_transposed}. MHL denotes the history-learning objective: its combination with OPQ is the MHL baseline, while SRBQ+MHL constitutes SAHL under the complete dataset-selected configuration. Under RPG, SRBQ improves NDCG@10 by 0.9\%--3.7\%. Adding history learning to SRBQ yields another 14.6\%--28.9\%, and SAHL leads every column. SRBQ+RPG changes how IDs and their units are constructed, whereas OPQ+MHL changes how histories are learned. Full SAHL connects the two by reconstructing the groups defined during SRBQ quantization.

\begin{figure}[!t]
\centering
\includegraphics[width=\linewidth]{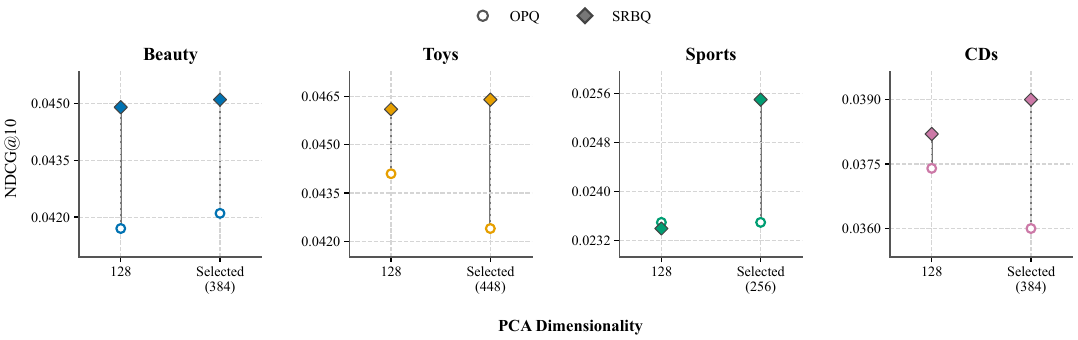}
\caption{Dimension-matched NDCG@10 comparison under direct next-item training. OPQ and SRBQ receive the same-dimensional PCA input and use the same semantic-ID length in each pair; circles and diamonds denote OPQ and SRBQ, respectively.}
\label{fig:dimension_matched}
\Description{Comparison of OPQ and SRBQ at matching embedding dimensions and semantic-ID lengths; circles and diamonds identify the quantizers.}
\end{figure}

\subsection{Dimension-Matched Quantizer Comparison}

Figure~\ref{fig:dimension_matched} controls retained dimensionality by comparing OPQ and SRBQ under direct training at 128 and the selected SRBQ dimension, with all other settings fixed within each pair. We find: at 128 dimensions, SRBQ outperforms others on three datasets. On Sports and Outdoors, its NDCG@10 is only .0001 lower (.0234 vs. .0235), whereas Recall@10 is .0004 higher (.0430 vs. .0426), making the isolated difference negligible. At the selected dimensions, SRBQ gains 7.1\%, 9.4\%, 8.5\%, and 8.3\% on Beauty, Toys and Games, Sports and Outdoors, and CDs and Vinyl, respectively, and outperforms others on seven of eight matched settings. The advantage persists under equal-dimensional inputs and cannot be attributed only to retaining more Sentence-T5 information.  Detailed results are given in Appendix~\ref{app:dimension_details}, Table~\ref{tab:app_dimension_matched}.
The 128-dimensional points reproduce OPQ's standard input budget, whereas the selected-dimensional points give both quantizers the same larger feature space. These comparisons attribute the gains to subspace residualization and balanced assignment, with competitive SRBQ performance under the standard 128-dimensional budget.

\begin{table*}[!t]
\centering
\caption{Multi-granularity masking under selected SRBQ, $\rho=0.15$, and R$\to$E$\to$Inf. Best results \textbf{bold}, second-best \underline{underlined}.}
\label{tab:srbq_mask_granularity}
\small
\begin{tabular*}{\linewidth}{@{\extracolsep{\fill}}lrrrrrrrr@{}}
\toprule
\textbf{Masking Strategy} & \multicolumn{4}{c}{\textbf{Beauty}} & \multicolumn{4}{c}{\textbf{Toys and Games}} \\
 & R@5 & N@5 & R@10 & N@10 & R@5 & N@5 & R@10 & N@10 \\
\midrule

Token & .0581 & \underline{.0425} & .0813 & \underline{.0500} & \underline{.0663} & \underline{.0471} & .0935 & \underline{.0558} \\
Item & .0550 & .0395 & .0779 & .0468 & .0643 & .0447 & .0875 & .0522 \\
Mixed & \underline{.0585} & .0413 & .0826 & .0491 & .0661 & .0453 & .0937 & .0541 \\
Subspace & \textbf{.0606} & \textbf{.0435} & \underline{.0831} & \textbf{.0507} & \textbf{.0696} & \textbf{.0491} & \textbf{.0962} & \textbf{.0577} \\
Tri-Mixed & .0571 & .0401 & \textbf{.0839} & .0487 & .0662 & .0466 & \underline{.0946} & .0557 \\
\bottomrule
\end{tabular*}
\par\medskip
\begin{tabular*}{\linewidth}{@{\extracolsep{\fill}}lrrrrrrrr@{}}
\toprule
\textbf{Masking Strategy} & \multicolumn{4}{c}{\textbf{Sports and Outdoors}} & \multicolumn{4}{c}{\textbf{CDs and Vinyl}} \\
 & R@5 & N@5 & R@10 & N@10 & R@5 & N@5 & R@10 & N@10 \\
\midrule

Token & \underline{.0369} & \underline{.0256} & \underline{.0533} & \underline{.0309} & .0540 & .0372 & .0779 & .0449 \\
Item & .0300 & .0206 & .0471 & .0260 & .0509 & .0350 & .0753 & .0428 \\
Mixed & .0360 & .0253 & \underline{.0533} & .0308 & \underline{.0545} & \underline{.0373} & \underline{.0798} & \underline{.0454} \\
Subspace & .0358 & .0238 & .0529 & .0294 & .0527 & .0363 & .0764 & .0439 \\
Tri-Mixed & \textbf{.0374} & \textbf{.0260} & \textbf{.0547} & \textbf{.0315} & \textbf{.0550} & \textbf{.0377} & \textbf{.0802} & \textbf{.0458} \\
\bottomrule
\end{tabular*}
\end{table*}

\subsection{Multi-Granularity Masking under SRBQ}
\label{sec:srbq_masking}

Table~\ref{tab:srbq_mask_granularity} compares five masks under selected SRBQ and fixed $\rho=0.15$. \textit{Entropy-Subspace} and \textit{Entropy-Tri-Mixed} reuse groups defined during SRBQ quantization. \textit{Entropy-Subspace} is best on Beauty and Toys and Games, exceeding the strongest OPQ-compatible mask by 1.4\%/3.4\%; \textit{Entropy-Tri-Mixed} is best on Sports and Outdoors and CDs and Vinyl, gaining 1.9\%/0.9\%. SRBQ-aligned reconstruction consistently helps, but the best granularity varies by catalog.

\begin{table}[!t]
\centering
\caption{Masking diagnostics for Beauty and for Toys and Games. Inf includes mask-ratio decay; probabilities follow \textnormal{\textit{Token}}/\textnormal{\textit{Subspace}}/\textnormal{\textit{Item}}. Best/second-best results are \textbf{bold}/\underline{underlined}.}
\label{tab:mask_granularity_diagnostic}
\renewcommand{\arraystretch}{1.16}
\begin{tabular*}{\linewidth}{@{\extracolsep{\fill}}>{\itshape}lcc@{}}
\toprule
\multirow{2}{*}{\textup{\textbf{Diagnostic Variant}}} & \textbf{Beauty} & \textbf{Toys and Games} \\
\cmidrule(lr){2-2}\cmidrule(lr){3-3}
& N@10 & N@10 \\
\midrule
\multicolumn{3}{c}{\textbf{Phase Attribution}} \\
\midrule
Subspace (R$\to$Inf) & \textbf{.0517} & \textbf{.0560} \\
Tri-Mixed (R$\to$Inf) & \underline{.0488} & .0472 \\
Subspace (E$\to$Inf) & .0474 & .0447 \\
Tri-Mixed (E$\to$Inf) & .0484 & \underline{.0548} \\
\midrule
\multicolumn{3}{c}{\textbf{Branch Attribution under R$\to$E$\to$Inf}} \\
\midrule
Token+Subspace (0.50/0.50/0.00) & \textbf{.0507} & \textbf{.0565} \\
Subspace+Item (0.00/0.50/0.50) & .0499 & .0537 \\
Subspace-heavy (0.15/0.70/0.15) & \underline{.0500} & \underline{.0554} \\
\bottomrule
\end{tabular*}
\end{table}

\subsection{Diagnostic Analysis of Masking Granularity}

Table~\ref{tab:mask_granularity_diagnostic} shows that under R$\to$Inf, \textit{Subspace} raises NDCG@10 over \textit{Tri-Mixed} from .0488 to .0517 on Beauty and .0472 to .0560 on Toys and Games. Under E$\to$Inf, \textit{Tri-Mixed} becomes competitive or stronger, locating its weakness before entropy stabilizes. With the full curriculum, equal \textit{Token}/\textit{Subspace} routing is best on both datasets; complete-item corruption is less useful than retaining partial token/subspace evidence.
Thus, the hierarchy offers alternative supervision units, while random warm-up controls unreliable early entropy.

\begin{table}[!t]
\centering
\caption{Structure-alignment ablation on Toys and Games. All variants use four-token reconstruction units, $\rho=0.15$, $\lambda_{\mathrm{rec}}=1.8$, and R$\to$E$\to$Inf. The two SRBQ rows keep the semantic IDs fixed. Best and second-best results are \textbf{bold} and \underline{underlined}.}
\label{tab:structure_alignment_toys}
\begin{tabular*}{\linewidth}{@{\extracolsep{\fill}}llccc@{}}
\toprule
\textbf{Quantizer} & \textbf{Grouping} & \textbf{Unit Size} & \textbf{R@10} & \textbf{N@10} \\
\midrule
OPQ & Pseudo-subspaces & 4 tokens & .0905 & .0562 \\
SRBQ & Randomly shuffled & 4 tokens & \underline{.0960} & \underline{.0584} \\
SRBQ & Aligned & 4 tokens & \textbf{.0964} & \textbf{.0598} \\
\bottomrule
\end{tabular*}
\end{table}

\subsection{Structure-Alignment Ablation}

Table~\ref{tab:structure_alignment_toys} distinguishes SRBQ structure from arbitrary blocks on Toys and Games. OPQ groups 32 digits into eight four-token pseudo-subspaces; SRBQ fixes IDs and unit size but uses aligned or randomly permuted groups.

OPQ pseudo-subspaces reach .0562 NDCG@10/.0905 Recall@10, below aligned SRBQ. Fixing SRBQ codes, unit size, and objective, using the quantizer-defined groups raises NDCG@10 from .0584 to .0598 (2.4\%) and Recall@10 from .0960 to .0964. This isolates grouping correspondence from ID values and block size.

The controls remove different alternatives: pseudo-subspaces test whether contiguous block corruption alone suffices, while shuffled SRBQ tests whether gains survive when the same tokens are grouped without latent correspondence. Improvement over both directly supports using the quantization structure as the supervision structure rather than merely introducing another block mask.

\begin{table}[!t]
\centering
\caption{History-dependence diagnostic on Toys and Games. The reconstruction target and evaluated samples are identical across conditions. Higher token accuracy and exact match are better; lower token NLL is better.}
\label{tab:history_dependence_toys}
\small
\renewcommand{\arraystretch}{1.08}
\begin{tabular*}{\linewidth}{@{\extracolsep{\fill}}lccc@{}}
\toprule
\textbf{Available Context} & \makecell[c]{\textbf{Token Acc.}\\$\uparrow$} & \makecell[c]{\textbf{Token NLL}\\$\downarrow$} & \makecell[c]{\textbf{Exact Match}\\$\uparrow$} \\
\midrule
\multicolumn{4}{c}{\textbf{History-Use Control: Same Current-Item Clues}} \\
\midrule
\makecell[l]{Full history +\\current clues} & .1739 & 3.687 & .0086 \\
\makecell[l]{Shuffled history +\\current clues} & .1548 & 3.882 & .0065 \\
\midrule
\multicolumn{4}{c}{\textbf{Evidence-Source Control}} \\
\midrule
\makecell[l]{Current clues only\\(no history)} & .2132 & 3.495 & .0174 \\
\makecell[l]{History only\\(no current clues)} & .0246 & 5.682 & .0015 \\
\bottomrule
\end{tabular*}
\end{table}

\subsection{History-Dependence Diagnostic}

Table~\ref{tab:history_dependence_toys} reconstructs a selected subspace under four contexts on Toys and Games to separate historical from current-item evidence. Holding the current item, visible tokens, and target fixed, shuffling the preceding history lowers token accuracy from .1739 to .1548, raises NLL from 3.687 to 3.882, and reduces exact match from .0086 to .0065. Current clues alone outperform history alone, showing that intra-item correlation is the easiest recovery cue; however, this condition cannot establish path modeling because semantic-ID tokens are correlated. Changing only the preceding user history consistently degrades all three metrics, confirming that subspace reconstruction also uses user-specific historical context rather than merely completing an item's internal code.

\begin{table*}[!t]
\centering
\caption{Case study on Beauty. The 15 most recent historical interactions are shown chronologically; italicized entries denote intervening non-hair-care interactions. Top-5 items follow predicted relevance, and the check mark denotes the ground-truth next item.}
\label{tab:case_study}
\footnotesize
\setlength{\tabcolsep}{4pt}
\renewcommand{\arraystretch}{1.08}
\begin{tabularx}{\linewidth}{>{\hsize=0.96\hsize\linewidth=\hsize\centering\arraybackslash}X|>{\hsize=0.96\hsize\linewidth=\hsize\centering\arraybackslash}X|>{\hsize=1.08\hsize\linewidth=\hsize\centering\arraybackslash}X}
\toprule
\multicolumn{3}{c}{\textbf{Historical Purchase Sequence}} \\
\midrule
\multicolumn{3}{p{0.97\textwidth}}{\centering
Motions Hydrate My Curls Pudding $\rightarrow$
Motions Define My Curls Creme $\rightarrow$
Remington Shine Therapy Flat Iron $\rightarrow$
\textit{Pond's Exfoliating Cleanser} $\rightarrow$
Motions Shine Enhancing Pomade $\rightarrow$
Motions Radiating Hair Gloss $\rightarrow$
Motions Moisturizing Cleanser $\rightarrow$
\textit{AINHOA Senskin Facial Concentrate} $\rightarrow$
Motions Deep Conditioning Masque $\rightarrow$
\textit{Simple Eye Make-Up Remover Pads} $\rightarrow$
\textit{L'Oreal RevitaLift Eye Treatment} $\rightarrow$
John Frieda Warm Blonde Hair Color $\rightarrow$
Fekkai PRX Oil $\rightarrow$
\textit{Hawaiian Tropic SPF 30 Face Lotion} $\rightarrow$
John Frieda Cool Brunette Hair Color $\rightarrow$ ?} \\
\midrule
\multicolumn{1}{c|}{\textbf{SAHL (Top-5)}} & \multicolumn{1}{c|}{\textbf{MHL (Top-5)}} & \multicolumn{1}{c}{\textbf{RPG (Top-5)}} \\
\midrule
John Frieda Cool Red Hair Color \textcolor{green!60!black}{\ding{51}} & John Frieda Warm Blonde Hair Color & TRESemm\'e Keratin Smooth Conditioner \\
Suave Gradual Self Tanner (Medium--Tan) & John Frieda Cool Blonde Hair Color & TRESemm\'e Keratin Smooth Shampoo \\
John Frieda Warm Red Hair Color & Suave Gradual Self Tanner (Medium--Tan) & Suave Gradual Self Tanner (Medium--Tan) \\
John Frieda Cool Blonde Hair Color & John Frieda Cool Red Hair Color \textcolor{green!60!black}{\ding{51}} & Caress Body Wash \\
John Frieda Warm Brunette Hair Color & John Frieda Cool Brunette Hair Color & Suave Gradual Self Tanner (Fair--Medium) \\
\bottomrule
\end{tabularx}
\end{table*}

\subsection{Further Analysis}

\subsubsection{Case Study}

Table~\ref{tab:case_study} shows a persistent hair-color trajectory interrupted by facial and sun-care items. RPG misses the Top-5 ground truth, returning two self-tanners and a body wash; its shampoo and conditioner capture only broad hair care. MHL returns four John Frieda colors but ranks the target fourth. SAHL ranks the target first with three same-brand, same-family alternatives, resolving the fine-grained preference. Although not causal evidence, this progression is consistent with history reconstruction and aligned IDs preserving persistent preferences under noise.

\begin{table}[!t]
\centering
\caption{Computational cost. Panel (a) measures history-learning training on Toys and Games; Panel (b) measures semantic-ID construction on Beauty. Both use $K=16$. Construction is timed once from cached Sentence-T5 embeddings, and relative time is computed against dimension-matched OPQ.}
\label{tab:computational_efficiency}
\small
\renewcommand{\arraystretch}{1.06}
\begin{tabular*}{\linewidth}{@{\extracolsep{\fill}}lrrr@{}}
\toprule
\multicolumn{4}{c}{\textbf{(a) History-Learning Training}} \\
\midrule
Method & Wall Clock (min) & Step (ms) & Memory (MB) \\
\midrule
RPG & 25.4 & 64.8 & 10,118 \\
MHL & 55.5 & 120.8 & 11,076 \\
\midrule
\multicolumn{4}{c}{\textbf{(b) Semantic-ID Construction}} \\
\midrule
Quantizer & PCA Dim. & Time (s) & vs. OPQ \\
\midrule
OPQ  & 128 & 62.5  & $1.00\times$ \\
SRBQ & 128 & 84.3  & $1.35\times$ \\
OPQ  & 384 & 79.0  & $1.00\times$ \\
SRBQ & 384 & 238.7 & $3.02\times$ \\
\bottomrule
\end{tabular*}
\end{table}

\subsubsection{Computational Efficiency}

Table~\ref{tab:computational_efficiency} uses one A6000 for matched RPG/MHL training with OPQ $K=16$, batch size 64, length 50, learning rate $5\times10^{-4}$, and FP16. MHL increases step/wall-clock time by 86.4\%/118.5\% and memory by 9.5\%; reconstruction is removed at inference, leaving RPG's decoder and graph propagation.

On Beauty, SRBQ takes 84.3/238.7 seconds at 128/384 dimensions versus 62.5/79.0 seconds for OPQ. This offline cost accompanies Fig.~\ref{fig:dimension_matched}'s accuracy gains.

Because IDs are constructed before recommender training and reused across requests, this overhead is amortized rather than paid per recommendation. At deployment, SAHL retains RPG's decoder architecture, beam size, and graph propagation, adding no online inference stage or cost.

\begin{table}[!t]
\centering
\caption{Recall@10 and NDCG@10 for transfer to item-title text tokens without a semantic-ID quantizer.}
\label{tab:text_token_transfer}
\begingroup
\small
\renewcommand{\arraystretch}{1.12}
\setlength{\tabcolsep}{3.0pt}
\begin{tabular*}{\linewidth}{@{\extracolsep{\fill}}lrrrrrrrr@{}}
\toprule
\multirow{2}{*}{\textbf{Training}} & \multicolumn{2}{c}{\textbf{Beauty}} & \multicolumn{2}{c}{\textbf{Toys and Games}} & \multicolumn{2}{c}{\textbf{Sports and Outdoors}} & \multicolumn{2}{c}{\textbf{CDs and Vinyl}} \\
\cmidrule(lr){2-3}\cmidrule(lr){4-5}\cmidrule(lr){6-7}\cmidrule(lr){8-9}
& \textbf{R@10} & \textbf{N@10} & \textbf{R@10} & \textbf{N@10} & \textbf{R@10} & \textbf{N@10} & \textbf{R@10} & \textbf{N@10} \\
\midrule
Inf & .0439 & .0258 & .0446 & .0273 & .0203 & .0117 & .0279 & .0162 \\
R$\to$E$\to$Inf & \textbf{.0483} & \textbf{.0285} & \textbf{.0498} & \textbf{.0297} & \textbf{.0237} & \textbf{.0134} & \textbf{.0343} & \textbf{.0189} \\
\bottomrule
\end{tabular*}
\endgroup
\end{table}

\begin{figure}[!t]
\centering
\includegraphics[width=\linewidth]{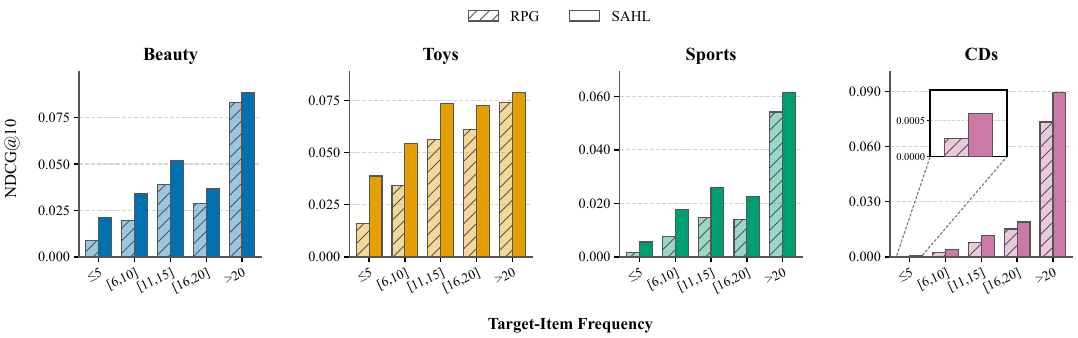}
\caption{RPG versus SAHL by target training frequency. The inset magnifies CDs and Vinyl at $\leq5$.}
\label{fig:cold_item_frequency}
\Description{RPG and SAHL ranking effectiveness across target-item training-frequency bins on four datasets, with an enlarged inset for rare CDs and Vinyl items.}
\end{figure}

\begin{figure}[!t]
\centering
\includegraphics[width=\linewidth]{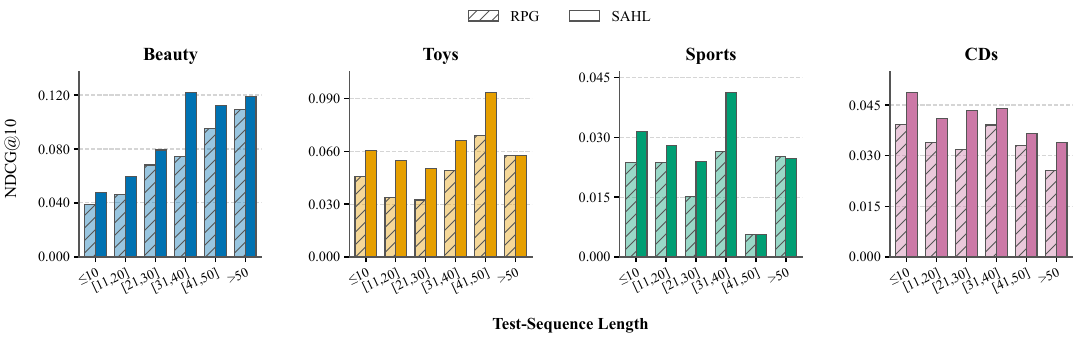}
\caption{RPG versus SAHL across test-history-length bins.}
\label{fig:sequence_length}
\Description{RPG and SAHL ranking effectiveness across test-history-length bins on four datasets.}
\end{figure}

\subsubsection{Transfer to Text-Token Prediction}

To test transfer beyond quantizers, we replace IDs with title-token sequences. Table~\ref{tab:text_token_transfer} compares direct training with R$\to$E$\to$Inf using \textit{Mixed} masking, $\rho=0.15$, and $\lambda_{\mathrm{rec}}=0.5$.

NDCG@10 rises by 8.8\%--16.7\% without a codebook. Title-token boundaries neither follow quantization positions nor expose subspace groups, so these gains are not a by-product of SRBQ assignments. They show that reconstruction, entropy guidance, and curriculum optimization provide a transferable history-learning mechanism, while the SRBQ experiments isolate the additional benefit of defining its reconstruction units during quantization.

\subsubsection{Low-Frequency (Cold-Item) Recommendation}

Following RPG~\cite{hou2025generating}, we group test targets by training frequency (near-cold, not zero-interaction). Figure~\ref{fig:cold_item_frequency} shows that SAHL wins every bin. At frequency $\leq5$, NDCG@10 improves by 100.0\%--205.6\% across datasets, versus 6.3\%--21.8\% above 20; gains are largest when item evidence is sparse. Detailed results are given in Appendix~\ref{app:robustness_details}, Table~\ref{tab:app_frequency}.

This pattern matches the intended sharing mechanism. Frequent targets already supervise complete IDs; rare targets depend more on reusable tokens and neighboring interactions. SAHL strengthens both sources.

\subsubsection{Effectiveness Across History Lengths}

Figure~\ref{fig:sequence_length} shows gains in every bin for Beauty and CDs and Vinyl and most bins elsewhere. Beauty improves throughout, and Toys and Games remains stronger through 41--50 interactions; Sports and Outdoors is nearly tied at 41--50 and reverses only marginally above 50. CDs and Vinyl rises from .0393 to .0487 at length $\leq10$ and .0257 to .0338 above 50. Thus, reconstruction helps short contexts; near ties in the longest bins suggest that rich histories partly offset weak supervision. Detailed results are given in Appendix~\ref{app:robustness_details}, Table~\ref{tab:app_history_length}.

\section{Conclusion}
In this work, we revisit generative recommendation from the perspective of history learning and propose SAHL to coordinate semantic-ID construction with historical supervision. SRBQ constructs a balanced hierarchy at the token, subspace, and item levels, which is directly reused to define multi-granularity reconstruction units along user histories. Entropy-guided target selection and curriculum optimization further encourage the model to capture persistent preferences while progressively aligning training with next-item inference. Extensive experiments demonstrate consistent improvements over strong generative recommendation baselines and confirm the complementary benefits of structured semantic IDs and explicit history supervision. Further analyses show that the gains stem from structure-aligned reconstruction and user-specific historical context rather than arbitrary token grouping. These findings indicate that semantic IDs can serve not only as generation targets, but also as structured interfaces for learning from historical behavior. Future work may explore adaptive tokenization, recommendation-aware assignment, and tighter joint optimization between semantic-ID construction and downstream recommendation.

\begin{acks}
This work was partially supported by the National Key R\&D Program of China (No. 2024YFC2707805), the National Natural Science Foundation of China (Nos. 62676401, 62676391, 62506050, 62406223, and 62176029), the Beijing Nova Program (No. 20250484899), the China Postdoctoral Science Foundation (No. 2024M763867), the Chongqing Science and Technology Innovation and Application Development Project (No. CSTB2023TIAD-KPX0064), the Open Competition Program of the Chongqing Municipal Commission of Economy and Information Technology (No. YJX-202500100200X), and the Chongqing Higher Education Teaching Reform Research Project (No. 242009).

\end{acks}




\clearpage
\appendix
\section{Detailed Experimental Results}
\label{app:experimental_details}
This appendix tabulates the experiments shown in Figures~\ref{fig:ablation_codebook_size}--\ref{fig:sequence_length}. R@5/R@10 and N@5/N@10 denote Recall and NDCG at cutoffs 5 and 10, respectively. All four metrics are reported. Each table follows the configuration and evaluation groups of its corresponding figure. All scores are rounded to four decimal places.

\subsection{OPQ Parameter Sensitivity}
\label{app:opq_details}
Tables~\ref{tab:app_opq_length}--\ref{tab:app_opq_loss} provide the numerical results for Figures~\ref{fig:ablation_codebook_size}--\ref{fig:ablation_recon_loss} under R$\to$E$\to$Inf.

\par\medskip
\noindent\begin{minipage}{\linewidth}
\centering
\captionsetup{hypcap=false}
\captionof{table}{Detailed results for Figure~\ref{fig:ablation_codebook_size}: OPQ semantic-ID length $K$, with $\rho=0.15$ for the first three datasets and $\rho=0.20$ for CDs and Vinyl; $\lambda_{\mathrm{rec}}=1.0$. $K\in\{4,8,16\}$.}
\label{tab:app_opq_length}
\begingroup
\small
\renewcommand{\arraystretch}{1.08}
\setlength{\tabcolsep}{2.0pt}
\begin{tabular*}{\linewidth}{@{\extracolsep{\fill}}llrrrrrrrr@{}}
\toprule
\textbf{$K$} & \textbf{Mask} & \multicolumn{4}{c}{\textbf{Beauty}} & \multicolumn{4}{c}{\textbf{Toys and Games}} \\
 &  & R@5 & N@5 & R@10 & N@10 & R@5 & N@5 & R@10 & N@10 \\
\midrule
4 & Token & .0406 & .0291 & .0568 & .0343 & .0434 & .0298 & .0638 & .0363 \\
4 & Item & .0344 & .0253 & .0516 & .0308 & .0385 & .0259 & .0576 & .0320 \\
4 & Mixed & .0376 & .0272 & .0531 & .0322 & .0422 & .0287 & .0603 & .0346 \\
\addlinespace[1pt]
8 & Token & .0502 & .0369 & .0691 & .0430 & .0577 & .0409 & .0805 & .0482 \\
8 & Item & .0458 & .0329 & .0646 & .0390 & .0503 & .0359 & .0716 & .0427 \\
8 & Mixed & .0468 & .0343 & .0645 & .0400 & .0553 & .0385 & .0785 & .0460 \\
\addlinespace[1pt]
16 & Token & .0574 & .0424 & .0795 & .0495 & .0644 & .0456 & .0883 & .0533 \\
16 & Item & .0501 & .0363 & .0704 & .0429 & .0553 & .0390 & .0802 & .0470 \\
16 & Mixed & .0537 & .0384 & .0757 & .0455 & .0592 & .0421 & .0845 & .0503 \\
\bottomrule
\end{tabular*}
\par\medskip
\begin{tabular*}{\linewidth}{@{\extracolsep{\fill}}llrrrrrrrr@{}}
\toprule
\textbf{$K$} & \textbf{Mask} & \multicolumn{4}{c}{\textbf{Sports and Outdoors}} & \multicolumn{4}{c}{\textbf{CDs and Vinyl}} \\
 &  & R@5 & N@5 & R@10 & N@10 & R@5 & N@5 & R@10 & N@10 \\
\midrule
4 & Token & .0177 & .0120 & .0277 & .0152 & .0220 & .0165 & .0298 & .0190 \\
4 & Item & .0151 & .0107 & .0242 & .0136 & .0203 & .0144 & .0288 & .0172 \\
4 & Mixed & .0203 & .0146 & .0302 & .0178 & .0206 & .0153 & .0293 & .0181 \\
\addlinespace[1pt]
8 & Token & .0280 & .0209 & .0395 & .0245 & .0337 & .0253 & .0448 & .0288 \\
8 & Item & .0249 & .0180 & .0366 & .0218 & .0362 & .0251 & .0518 & .0301 \\
8 & Mixed & .0279 & .0201 & .0389 & .0237 & .0369 & .0265 & .0514 & .0311 \\
\addlinespace[1pt]
16 & Token & .0334 & .0239 & .0474 & .0284 & .0470 & .0330 & .0672 & .0395 \\
16 & Item & .0282 & .0197 & .0419 & .0241 & .0440 & .0301 & .0637 & .0364 \\
16 & Mixed & .0325 & .0228 & .0478 & .0278 & .0475 & .0330 & .0674 & .0394 \\
\bottomrule
\end{tabular*}
\endgroup
\end{minipage}\par\medskip

\par\medskip
\noindent\begin{minipage}{\linewidth}
\centering
\captionsetup{hypcap=false}
\captionsetup{type=table}
\ContinuedFloat
\caption[]{Detailed results for Figure~\ref{fig:ablation_codebook_size}: OPQ semantic-ID length $K$, with $\rho=0.15$ for the first three datasets and $\rho=0.20$ for CDs and Vinyl; $\lambda_{\mathrm{rec}}=1.0$. $K\in\{32,64\}$. \textit{Continued.}}
\begingroup
\small
\renewcommand{\arraystretch}{1.08}
\setlength{\tabcolsep}{2.0pt}
\begin{tabular*}{\linewidth}{@{\extracolsep{\fill}}llrrrrrrrr@{}}
\toprule
\textbf{$K$} & \textbf{Mask} & \multicolumn{4}{c}{\textbf{Beauty}} & \multicolumn{4}{c}{\textbf{Toys and Games}} \\
 &  & R@5 & N@5 & R@10 & N@10 & R@5 & N@5 & R@10 & N@10 \\
\midrule
32 & Token & .0568 & .0411 & .0814 & .0490 & .0634 & .0458 & .0880 & .0538 \\
32 & Item & .0523 & .0368 & .0743 & .0439 & .0579 & .0416 & .0811 & .0491 \\
32 & Mixed & .0537 & .0383 & .0784 & .0463 & .0613 & .0434 & .0855 & .0511 \\
\addlinespace[1pt]
64 & Token & .0552 & .0390 & .0805 & .0472 & .0618 & .0448 & .0861 & .0527 \\
64 & Item & .0509 & .0358 & .0731 & .0429 & .0546 & .0384 & .0768 & .0456 \\
64 & Mixed & .0533 & .0377 & .0760 & .0451 & .0579 & .0411 & .0824 & .0490 \\
\bottomrule
\end{tabular*}
\par\medskip
\begin{tabular*}{\linewidth}{@{\extracolsep{\fill}}llrrrrrrrr@{}}
\toprule
\textbf{$K$} & \textbf{Mask} & \multicolumn{4}{c}{\textbf{Sports and Outdoors}} & \multicolumn{4}{c}{\textbf{CDs and Vinyl}} \\
 &  & R@5 & N@5 & R@10 & N@10 & R@5 & N@5 & R@10 & N@10 \\
\midrule
32 & Token & .0344 & .0244 & .0495 & .0292 & .0482 & .0336 & .0689 & .0403 \\
32 & Item & .0286 & .0193 & .0421 & .0236 & .0485 & .0332 & .0710 & .0405 \\
32 & Mixed & .0334 & .0235 & .0489 & .0285 & .0505 & .0347 & .0729 & .0419 \\
\addlinespace[1pt]
64 & Token & .0325 & .0226 & .0475 & .0275 & .0487 & .0339 & .0707 & .0409 \\
64 & Item & .0273 & .0179 & .0417 & .0226 & .0491 & .0334 & .0723 & .0409 \\
64 & Mixed & .0318 & .0218 & .0471 & .0267 & .0526 & .0363 & .0768 & .0441 \\
\bottomrule
\end{tabular*}
\endgroup
\end{minipage}\par\medskip

\par\medskip
\noindent\begin{minipage}{\linewidth}
\centering
\captionsetup{hypcap=false}
\captionof{table}{Detailed results for Figure~\ref{fig:ablation_mask_ratio}: OPQ masking ratio $\rho$, with $K=16/32/32/64$ in dataset order and $\lambda_{\mathrm{rec}}=1.0$. $\rho\in\{0.05,0.10,0.15,0.20\}$.}
\label{tab:app_opq_ratio}
\begingroup
\small
\renewcommand{\arraystretch}{1.08}
\setlength{\tabcolsep}{2.0pt}
\begin{tabular*}{\linewidth}{@{\extracolsep{\fill}}llrrrrrrrr@{}}
\toprule
\textbf{$\rho$} & \textbf{Mask} & \multicolumn{4}{c}{\textbf{Beauty}} & \multicolumn{4}{c}{\textbf{Toys and Games}} \\
 &  & R@5 & N@5 & R@10 & N@10 & R@5 & N@5 & R@10 & N@10 \\
\midrule
0.05 & Token & .0562 & .0410 & .0782 & .0481 & .0657 & .0476 & .0898 & .0553 \\
0.05 & Item & .0327 & .0225 & .0499 & .0280 & .0565 & .0400 & .0778 & .0469 \\
0.05 & Mixed & .0544 & .0392 & .0758 & .0461 & .0596 & .0426 & .0821 & .0498 \\
\addlinespace[1pt]
0.10 & Token & .0572 & .0412 & .0805 & .0487 & .0632 & .0463 & .0885 & .0544 \\
0.10 & Item & .0495 & .0359 & .0700 & .0424 & .0583 & .0420 & .0810 & .0492 \\
0.10 & Mixed & .0548 & .0397 & .0767 & .0467 & .0610 & .0440 & .0855 & .0519 \\
\addlinespace[1pt]
0.15 & Token & .0574 & .0424 & .0795 & .0495 & .0634 & .0458 & .0880 & .0538 \\
0.15 & Item & .0501 & .0363 & .0704 & .0429 & .0579 & .0416 & .0811 & .0491 \\
0.15 & Mixed & .0537 & .0384 & .0757 & .0455 & .0613 & .0434 & .0855 & .0511 \\
\addlinespace[1pt]
0.20 & Token & .0549 & .0403 & .0788 & .0480 & .0628 & .0457 & .0858 & .0531 \\
0.20 & Item & .0507 & .0372 & .0716 & .0439 & .0568 & .0408 & .0819 & .0489 \\
0.20 & Mixed & .0560 & .0401 & .0775 & .0470 & .0617 & .0440 & .0853 & .0516 \\
\bottomrule
\end{tabular*}
\par\medskip
\begin{tabular*}{\linewidth}{@{\extracolsep{\fill}}llrrrrrrrr@{}}
\toprule
\textbf{$\rho$} & \textbf{Mask} & \multicolumn{4}{c}{\textbf{Sports and Outdoors}} & \multicolumn{4}{c}{\textbf{CDs and Vinyl}} \\
 &  & R@5 & N@5 & R@10 & N@10 & R@5 & N@5 & R@10 & N@10 \\
\midrule
0.05 & Token & .0341 & .0236 & .0499 & .0287 & .0488 & .0341 & .0717 & .0415 \\
0.05 & Item & .0277 & .0183 & .0418 & .0229 & .0470 & .0321 & .0684 & .0390 \\
0.05 & Mixed & .0323 & .0225 & .0485 & .0277 & .0516 & .0356 & .0758 & .0434 \\
\addlinespace[1pt]
0.10 & Token & .0346 & .0246 & .0493 & .0293 & .0481 & .0334 & .0692 & .0402 \\
0.10 & Item & .0301 & .0201 & .0433 & .0244 & .0491 & .0336 & .0725 & .0411 \\
0.10 & Mixed & .0342 & .0236 & .0483 & .0280 & .0516 & .0352 & .0747 & .0427 \\
\addlinespace[1pt]
0.15 & Token & .0344 & .0244 & .0495 & .0292 & .0369 & .0254 & .0529 & .0306 \\
0.15 & Item & .0286 & .0193 & .0421 & .0236 & .0489 & .0334 & .0726 & .0410 \\
0.15 & Mixed & .0334 & .0235 & .0489 & .0285 & .0501 & .0345 & .0730 & .0418 \\
\addlinespace[1pt]
0.20 & Token & .0339 & .0240 & .0473 & .0283 & .0487 & .0339 & .0707 & .0409 \\
0.20 & Item & .0290 & .0195 & .0426 & .0239 & .0491 & .0334 & .0723 & .0409 \\
0.20 & Mixed & .0342 & .0238 & .0490 & .0286 & .0526 & .0363 & .0768 & .0441 \\
\bottomrule
\end{tabular*}
\endgroup
\end{minipage}\par\medskip

\par\medskip
\noindent\begin{minipage}{\linewidth}
\centering
\captionsetup{hypcap=false}
\captionsetup{type=table}
\ContinuedFloat
\caption[]{Detailed results for Figure~\ref{fig:ablation_mask_ratio}: OPQ masking ratio $\rho$, with $K=16/32/32/64$ in dataset order and $\lambda_{\mathrm{rec}}=1.0$. $\rho\in\{0.25,0.30,0.35,0.40\}$. \textit{Continued.}}
\begingroup
\small
\renewcommand{\arraystretch}{1.08}
\setlength{\tabcolsep}{2.0pt}
\begin{tabular*}{\linewidth}{@{\extracolsep{\fill}}llrrrrrrrr@{}}
\toprule
\textbf{$\rho$} & \textbf{Mask} & \multicolumn{4}{c}{\textbf{Beauty}} & \multicolumn{4}{c}{\textbf{Toys and Games}} \\
 &  & R@5 & N@5 & R@10 & N@10 & R@5 & N@5 & R@10 & N@10 \\
\midrule
0.25 & Token & .0563 & .0409 & .0795 & .0484 & .0624 & .0447 & .0873 & .0527 \\
0.25 & Item & .0511 & .0372 & .0717 & .0439 & .0568 & .0403 & .0800 & .0477 \\
0.25 & Mixed & .0529 & .0382 & .0746 & .0452 & .0617 & .0433 & .0860 & .0512 \\
\addlinespace[1pt]
0.30 & Token & .0558 & .0407 & .0774 & .0476 & .0613 & .0443 & .0854 & .0521 \\
0.30 & Item & .0510 & .0370 & .0715 & .0436 & .0580 & .0404 & .0815 & .0479 \\
0.30 & Mixed & .0519 & .0377 & .0743 & .0449 & .0612 & .0428 & .0854 & .0506 \\
\addlinespace[1pt]
0.35 & Token & .0532 & .0389 & .0767 & .0465 & .0615 & .0439 & .0841 & .0512 \\
0.35 & Item & .0521 & .0379 & .0749 & .0453 & .0571 & .0395 & .0807 & .0471 \\
0.35 & Mixed & .0517 & .0372 & .0757 & .0449 & .0598 & .0422 & .0848 & .0503 \\
\addlinespace[1pt]
0.40 & Token & .0551 & .0399 & .0772 & .0471 & .0593 & .0429 & .0844 & .0510 \\
0.40 & Item & .0516 & .0370 & .0740 & .0442 & .0555 & .0384 & .0806 & .0466 \\
0.40 & Mixed & .0516 & .0368 & .0753 & .0444 & .0590 & .0418 & .0825 & .0494 \\
\bottomrule
\end{tabular*}
\par\medskip
\begin{tabular*}{\linewidth}{@{\extracolsep{\fill}}llrrrrrrrr@{}}
\toprule
\textbf{$\rho$} & \textbf{Mask} & \multicolumn{4}{c}{\textbf{Sports and Outdoors}} & \multicolumn{4}{c}{\textbf{CDs and Vinyl}} \\
 &  & R@5 & N@5 & R@10 & N@10 & R@5 & N@5 & R@10 & N@10 \\
\midrule
0.25 & Token & .0339 & .0239 & .0467 & .0280 & .0490 & .0341 & .0717 & .0414 \\
0.25 & Item & .0292 & .0193 & .0439 & .0241 & .0495 & .0339 & .0722 & .0411 \\
0.25 & Mixed & .0335 & .0235 & .0480 & .0282 & .0530 & .0363 & .0767 & .0439 \\
\addlinespace[1pt]
0.30 & Token & .0333 & .0238 & .0466 & .0281 & .0490 & .0336 & .0713 & .0408 \\
0.30 & Item & .0295 & .0200 & .0447 & .0248 & .0495 & .0336 & .0727 & .0411 \\
0.30 & Mixed & .0335 & .0232 & .0489 & .0282 & .0522 & .0359 & .0761 & .0437 \\
\addlinespace[1pt]
0.35 & Token & .0333 & .0236 & .0472 & .0280 & .0507 & .0347 & .0726 & .0417 \\
0.35 & Item & .0285 & .0192 & .0446 & .0243 & .0505 & .0345 & .0747 & .0423 \\
0.35 & Mixed & .0340 & .0237 & .0490 & .0285 & .0520 & .0357 & .0758 & .0433 \\
\addlinespace[1pt]
0.40 & Token & .0336 & .0237 & .0474 & .0282 & .0507 & .0347 & .0718 & .0415 \\
0.40 & Item & .0296 & .0199 & .0439 & .0245 & .0505 & .0343 & .0739 & .0418 \\
0.40 & Mixed & .0329 & .0228 & .0490 & .0280 & .0505 & .0348 & .0741 & .0424 \\
\bottomrule
\end{tabular*}
\endgroup
\end{minipage}\par\medskip

\par\medskip
\noindent\begin{minipage}{\linewidth}
\centering
\captionsetup{hypcap=false}
\captionof{table}{Detailed results for Figure~\ref{fig:ablation_recon_loss}: reconstruction weight $\lambda_{\mathrm{rec}}$ under the selected OPQ length and masking configuration; prediction weight is 1.0.}
\label{tab:app_opq_loss}
\begingroup
\small
\renewcommand{\arraystretch}{1.08}
\setlength{\tabcolsep}{2.0pt}
\begin{tabular*}{\linewidth}{@{\extracolsep{\fill}}lrrrrrrrr@{}}
\toprule
\textbf{$\lambda_{\mathrm{rec}}$} & \multicolumn{4}{c}{\textbf{Beauty}} & \multicolumn{4}{c}{\textbf{Toys and Games}} \\
 & R@5 & N@5 & R@10 & N@10 & R@5 & N@5 & R@10 & N@10 \\
\midrule
0.2 & .0526 & .0376 & .0758 & .0450 & .0602 & .0429 & .0854 & .0510 \\
0.4 & .0533 & .0382 & .0768 & .0458 & .0611 & .0435 & .0856 & .0513 \\
0.6 & .0549 & .0401 & .0790 & .0479 & .0640 & .0460 & .0881 & .0537 \\
0.8 & .0554 & .0404 & .0793 & .0481 & .0642 & .0458 & .0890 & .0538 \\
1.0 & .0574 & .0424 & .0795 & .0495 & .0657 & .0476 & .0898 & .0553 \\
1.2 & .0559 & .0412 & .0774 & .0481 & .0646 & .0474 & .0867 & .0545 \\
1.4 & .0559 & .0413 & .0800 & .0490 & .0664 & .0489 & .0896 & .0563 \\
1.6 & .0564 & .0415 & .0767 & .0480 & .0653 & .0476 & .0877 & .0548 \\
1.8 & .0560 & .0411 & .0769 & .0479 & .0672 & .0489 & .0903 & .0564 \\
2.0 & .0562 & .0414 & .0779 & .0484 & .0638 & .0475 & .0868 & .0548 \\
\bottomrule
\end{tabular*}
\par\medskip
\begin{tabular*}{\linewidth}{@{\extracolsep{\fill}}lrrrrrrrr@{}}
\toprule
\textbf{$\lambda_{\mathrm{rec}}$} & \multicolumn{4}{c}{\textbf{Sports and Outdoors}} & \multicolumn{4}{c}{\textbf{CDs and Vinyl}} \\
 & R@5 & N@5 & R@10 & N@10 & R@5 & N@5 & R@10 & N@10 \\
\midrule
0.2 & .0330 & .0224 & .0487 & .0275 & .0529 & .0357 & .0776 & .0436 \\
0.4 & .0330 & .0235 & .0465 & .0278 & .0530 & .0358 & .0780 & .0438 \\
0.6 & .0331 & .0235 & .0478 & .0283 & .0539 & .0369 & .0787 & .0449 \\
0.8 & .0348 & .0244 & .0482 & .0287 & .0530 & .0365 & .0766 & .0441 \\
1.0 & .0346 & .0246 & .0493 & .0293 & .0526 & .0363 & .0768 & .0441 \\
1.2 & .0359 & .0249 & .0511 & .0298 & .0507 & .0350 & .0740 & .0425 \\
1.4 & .0361 & .0253 & .0503 & .0298 & .0508 & .0349 & .0737 & .0422 \\
1.6 & .0344 & .0241 & .0484 & .0286 & .0454 & .0312 & .0656 & .0377 \\
1.8 & .0347 & .0246 & .0484 & .0290 & .0498 & .0342 & .0719 & .0413 \\
2.0 & .0337 & .0241 & .0472 & .0285 & .0506 & .0348 & .0741 & .0423 \\
\bottomrule
\end{tabular*}
\endgroup
\end{minipage}\par\medskip

\clearpage
\subsection{SRBQ Parameter Sensitivity}
\label{app:srbq_details}
Tables~\ref{tab:app_srbq_dimension}--\ref{tab:app_srbq_balance} give the three sequential searches in Figure~\ref{fig:srbq_parameter_search}. Bold entries mark the configurations selected in that figure, rather than independently maximizing each metric. Step~1 fixes $G=4/8/8/16$ and $\beta=1.5$. Step~2 fixes the selected dimensions $384/448/256/384$ and $\beta=1.5$. Step~3 fixes those dimensions and the selected $G=4/8/8/16$, all in dataset order.

\par\medskip
\noindent\begin{minipage}{\linewidth}
\centering
\captionsetup{hypcap=false}
\captionof{table}{Detailed results for Figure~\ref{fig:srbq_parameter_search}, Step~1: PCA dimension $d_{\mathrm{srbq}}$. Bold entries mark the selected setting for each dataset.}
\label{tab:app_srbq_dimension}
\begingroup
\small
\renewcommand{\arraystretch}{1.08}
\setlength{\tabcolsep}{2.0pt}
\begin{tabular*}{\linewidth}{@{\extracolsep{\fill}}lrrrrrrrr@{}}
\toprule
\textbf{$d_{\mathrm{srbq}}$} & \multicolumn{4}{c}{\textbf{Beauty}} & \multicolumn{4}{c}{\textbf{Toys and Games}} \\
 & R@5 & N@5 & R@10 & N@10 & R@5 & N@5 & R@10 & N@10 \\
\midrule
128 & .0599 & .0437 & .0839 & .0514 & .0684 & .0496 & .0943 & .0579 \\
256 & .0600 & .0428 & .0818 & .0498 & .0710 & .0514 & .0962 & .0594 \\
384 & \textbf{.0601} & \textbf{.0432} & \textbf{.0861} & \textbf{.0516} & .0669 & .0480 & .0944 & .0568 \\
448 & .0592 & .0432 & .0842 & .0512 & \textbf{.0711} & \textbf{.0508} & \textbf{.0977} & \textbf{.0594} \\
512 & .0593 & .0423 & .0849 & .0506 & .0684 & .0488 & .0932 & .0568 \\
\bottomrule
\end{tabular*}
\par\medskip
\begin{tabular*}{\linewidth}{@{\extracolsep{\fill}}lrrrrrrrr@{}}
\toprule
\textbf{$d_{\mathrm{srbq}}$} & \multicolumn{4}{c}{\textbf{Sports and Outdoors}} & \multicolumn{4}{c}{\textbf{CDs and Vinyl}} \\
 & R@5 & N@5 & R@10 & N@10 & R@5 & N@5 & R@10 & N@10 \\
\midrule
128 & .0353 & .0251 & .0500 & .0298 & .0539 & .0371 & .0793 & .0453 \\
256 & \textbf{.0367} & \textbf{.0256} & \textbf{.0546} & \textbf{.0313} & .0542 & .0372 & .0794 & .0453 \\
384 & .0361 & .0257 & .0515 & .0306 & \textbf{.0550} & \textbf{.0373} & \textbf{.0807} & \textbf{.0456} \\
448 & .0369 & .0255 & .0542 & .0310 & .0536 & .0363 & .0793 & .0445 \\
512 & .0360 & .0250 & .0537 & .0307 & .0542 & .0369 & .0803 & .0453 \\
\bottomrule
\end{tabular*}
\endgroup
\end{minipage}\par\medskip

\par\medskip
\noindent\begin{minipage}{\linewidth}
\centering
\captionsetup{hypcap=false}
\captionof{table}{Figure~\ref{fig:srbq_parameter_search}, Step~2: subspace-count results. Bold marks selected settings; a dash denotes $G>K$.}
\label{tab:app_srbq_groups}
\begingroup
\small
\renewcommand{\arraystretch}{1.08}
\setlength{\tabcolsep}{2.0pt}
\begin{tabular*}{\linewidth}{@{\extracolsep{\fill}}lrrrrrrrr@{}}
\toprule
\textbf{$G$} & \multicolumn{4}{c}{\textbf{Beauty}} & \multicolumn{4}{c}{\textbf{Toys and Games}} \\
 & R@5 & N@5 & R@10 & N@10 & R@5 & N@5 & R@10 & N@10 \\
\midrule
2 & .0572 & .0422 & .0786 & .0491 & .0652 & .0473 & .0870 & .0543 \\
4 & \textbf{.0601} & \textbf{.0432} & \textbf{.0861} & \textbf{.0516} & .0698 & .0510 & .0928 & .0584 \\
8 & .0595 & .0425 & .0854 & .0509 & \textbf{.0711} & \textbf{.0508} & \textbf{.0977} & \textbf{.0594} \\
16 & .0579 & .0412 & .0804 & .0484 & .0704 & .0487 & .0988 & .0579 \\
32 & \textemdash & \textemdash & \textemdash & \textemdash & .0686 & .0493 & .0955 & .0580 \\
\bottomrule
\end{tabular*}
\par\medskip
\begin{tabular*}{\linewidth}{@{\extracolsep{\fill}}lrrrrrrrr@{}}
\toprule
\textbf{$G$} & \multicolumn{4}{c}{\textbf{Sports and Outdoors}} & \multicolumn{4}{c}{\textbf{CDs and Vinyl}} \\
 & R@5 & N@5 & R@10 & N@10 & R@5 & N@5 & R@10 & N@10 \\
\midrule
2 & .0343 & .0242 & .0478 & .0285 & .0504 & .0341 & .0762 & .0429 \\
4 & .0358 & .0255 & .0507 & .0303 & .0518 & .0354 & .0765 & .0434 \\
8 & \textbf{.0367} & \textbf{.0256} & \textbf{.0546} & \textbf{.0313} & .0541 & .0365 & .0797 & .0447 \\
16 & .0369 & .0260 & .0537 & .0313 & \textbf{.0550} & \textbf{.0373} & \textbf{.0807} & \textbf{.0456} \\
32 & .0351 & .0249 & .0507 & .0299 & .0539 & .0368 & .0791 & .0449 \\
\bottomrule
\end{tabular*}
\endgroup
\end{minipage}\par\medskip

\par\medskip
\noindent\begin{minipage}{\linewidth}
\centering
\captionsetup{hypcap=false}
\captionof{table}{Detailed results for Figure~\ref{fig:srbq_parameter_search}, Step~3: balance ratio $\beta$. Bold entries mark the selected setting for each dataset.}
\label{tab:app_srbq_balance}
\begingroup
\small
\renewcommand{\arraystretch}{1.08}
\setlength{\tabcolsep}{2.0pt}
\begin{tabular*}{\linewidth}{@{\extracolsep{\fill}}lrrrrrrrr@{}}
\toprule
\textbf{$\beta$} & \multicolumn{4}{c}{\textbf{Beauty}} & \multicolumn{4}{c}{\textbf{Toys and Games}} \\
 & R@5 & N@5 & R@10 & N@10 & R@5 & N@5 & R@10 & N@10 \\
\midrule
1.0 & .0590 & .0427 & .0853 & .0512 & .0686 & .0500 & .0947 & .0584 \\
1.2 & .0589 & .0429 & .0835 & .0508 & .0687 & .0502 & .0940 & .0584 \\
1.5 & \textbf{.0601} & \textbf{.0432} & \textbf{.0861} & \textbf{.0516} & \textbf{.0711} & \textbf{.0508} & \textbf{.0977} & \textbf{.0594} \\
2.0 & .0592 & .0429 & .0833 & .0507 & .0689 & .0490 & .0950 & .0575 \\
3.0 & .0591 & .0432 & .0814 & .0503 & .0706 & .0502 & .0960 & .0584 \\
\bottomrule
\end{tabular*}
\par\medskip
\begin{tabular*}{\linewidth}{@{\extracolsep{\fill}}lrrrrrrrr@{}}
\toprule
\textbf{$\beta$} & \multicolumn{4}{c}{\textbf{Sports and Outdoors}} & \multicolumn{4}{c}{\textbf{CDs and Vinyl}} \\
 & R@5 & N@5 & R@10 & N@10 & R@5 & N@5 & R@10 & N@10 \\
\midrule
1.0 & .0351 & .0247 & .0512 & .0299 & .0552 & .0379 & .0800 & .0462 \\
1.2 & .0373 & .0254 & .0544 & .0312 & \textbf{.0560} & \textbf{.0380} & \textbf{.0822} & \textbf{.0464} \\
1.5 & \textbf{.0367} & \textbf{.0256} & \textbf{.0546} & \textbf{.0313} & .0550 & .0373 & .0807 & .0456 \\
2.0 & .0367 & .0249 & .0542 & .0309 & .0545 & .0371 & .0792 & .0450 \\
3.0 & .0370 & .0259 & .0517 & .0307 & .0548 & .0371 & .0797 & .0452 \\
\bottomrule
\end{tabular*}
\endgroup
\end{minipage}\par\medskip

\subsection{Dimension-Matched Quantizer Comparison}
\label{app:dimension_details}
Table~\ref{tab:app_dimension_matched} gives the direct-training results in Figure~\ref{fig:dimension_matched}. Both quantizers use 128 dimensions in the first pair; the selected-dimensional pair uses 384, 448, 256, and 384 dimensions on Beauty, Toys and Games, Sports and Outdoors, and CDs and Vinyl, respectively.

\par\medskip
\noindent\begin{minipage}{\linewidth}
\centering
\captionsetup{hypcap=false}
\captionof{table}{Detailed results for Figure~\ref{fig:dimension_matched}: OPQ and SRBQ under direct next-item training with matching PCA dimensionality.}
\label{tab:app_dimension_matched}
\begingroup
\small
\renewcommand{\arraystretch}{1.08}
\setlength{\tabcolsep}{2.0pt}
\begin{tabular*}{\linewidth}{@{\extracolsep{\fill}}llrrrrrrrr@{}}
\toprule
\textbf{Dimension} & \textbf{Method} & \multicolumn{4}{c}{\textbf{Beauty}} & \multicolumn{4}{c}{\textbf{Toys and Games}} \\
 &  & R@5 & N@5 & R@10 & N@10 & R@5 & N@5 & R@10 & N@10 \\
\midrule
128 & OPQ & .0493 & .0348 & .0710 & .0417 & .0524 & .0367 & .0756 & .0441 \\
128 & SRBQ & .0524 & .0367 & .0776 & .0449 & .0556 & .0386 & .0789 & .0461 \\
\addlinespace[1pt]
Selected & OPQ & .0485 & .0346 & .0716 & .0421 & .0511 & .0347 & .0751 & .0424 \\
Selected & SRBQ & .0536 & .0367 & .0796 & .0451 & .0555 & .0377 & .0825 & .0464 \\
\bottomrule
\end{tabular*}
\par\medskip
\begin{tabular*}{\linewidth}{@{\extracolsep{\fill}}llrrrrrrrr@{}}
\toprule
\textbf{Dimension} & \textbf{Method} & \multicolumn{4}{c}{\textbf{Sports and Outdoors}} & \multicolumn{4}{c}{\textbf{CDs and Vinyl}} \\
 &  & R@5 & N@5 & R@10 & N@10 & R@5 & N@5 & R@10 & N@10 \\
\midrule
128 & OPQ & .0286 & .0190 & .0426 & .0235 & .0455 & .0305 & .0669 & .0374 \\
128 & SRBQ & .0282 & .0187 & .0430 & .0234 & .0464 & .0308 & .0695 & .0382 \\
\addlinespace[1pt]
Selected & OPQ & .0275 & .0188 & .0419 & .0235 & .0432 & .0296 & .0631 & .0360 \\
Selected & SRBQ & .0296 & .0204 & .0455 & .0255 & .0472 & .0318 & .0696 & .0390 \\
\bottomrule
\end{tabular*}
\endgroup
\end{minipage}\par\medskip

\clearpage
\subsection{Target Frequency and History Length}
\label{app:robustness_details}
Tables~\ref{tab:app_frequency} and~\ref{tab:app_history_length} provide the grouped results in Figures~\ref{fig:cold_item_frequency} and~\ref{fig:sequence_length}. Frequency counts refer to target-item occurrences in the training set; history-length bins refer to the test user's interaction history.

\par\medskip
\noindent\begin{minipage}{\linewidth}
\centering
\captionsetup{hypcap=false}
\captionof{table}{Detailed results for Figure~\ref{fig:cold_item_frequency}: target-item training-frequency groups.}
\label{tab:app_frequency}
\begingroup
\small
\renewcommand{\arraystretch}{1.08}
\setlength{\tabcolsep}{2.0pt}
\begin{tabular*}{\linewidth}{@{\extracolsep{\fill}}llrrrrrrrr@{}}
\toprule
\textbf{Frequency} & \textbf{Method} & \multicolumn{4}{c}{\textbf{Beauty}} & \multicolumn{4}{c}{\textbf{Toys and Games}} \\
 &  & R@5 & N@5 & R@10 & N@10 & R@5 & N@5 & R@10 & N@10 \\
\midrule
$\leq 5$ & RPG & .0084 & .0049 & .0203 & .0086 & .0184 & .0116 & .0323 & .0161 \\
$\leq 5$ & SAHL & .0247 & .0165 & .0391 & .0211 & .0456 & .0318 & .0672 & .0387 \\
\addlinespace[1pt]
$[6,10]$ & RPG & .0249 & .0157 & .0371 & .0197 & .0401 & .0262 & .0656 & .0343 \\
$[6,10]$ & SAHL & .0418 & .0293 & .0570 & .0342 & .0661 & .0454 & .0932 & .0542 \\
\addlinespace[1pt]
$[11,15]$ & RPG & .0453 & .0331 & .0630 & .0388 & .0683 & .0480 & .0943 & .0564 \\
$[11,15]$ & SAHL & .0583 & .0438 & .0831 & .0518 & .0822 & .0636 & .1137 & .0737 \\
\addlinespace[1pt]
$[16,20]$ & RPG & .0327 & .0224 & .0523 & .0288 & .0702 & .0519 & .0983 & .0610 \\
$[16,20]$ & SAHL & .0445 & .0317 & .0595 & .0365 & .0860 & .0641 & .1128 & .0727 \\
\addlinespace[1pt]
$>20$ & RPG & .0994 & .0716 & .1352 & .0832 & .0892 & .0644 & .1186 & .0738 \\
$>20$ & SAHL & .1028 & .0759 & .1416 & .0884 & .0927 & .0698 & .1206 & .0787 \\
\bottomrule
\end{tabular*}
\par\medskip
\begin{tabular*}{\linewidth}{@{\extracolsep{\fill}}llrrrrrrrr@{}}
\toprule
\textbf{Frequency} & \textbf{Method} & \multicolumn{4}{c}{\textbf{Sports and Outdoors}} & \multicolumn{4}{c}{\textbf{CDs and Vinyl}} \\
 &  & R@5 & N@5 & R@10 & N@10 & R@5 & N@5 & R@10 & N@10 \\
\midrule
$\leq 5$ & RPG & .0018 & .0008 & .0047 & .0018 & .0003 & .0001 & .0007 & .0003 \\
$\leq 5$ & SAHL & .0057 & .0033 & .0128 & .0055 & .0008 & .0003 & .0016 & .0006 \\
\addlinespace[1pt]
$[6,10]$ & RPG & .0094 & .0056 & .0154 & .0076 & .0029 & .0015 & .0057 & .0024 \\
$[6,10]$ & SAHL & .0213 & .0143 & .0321 & .0178 & .0038 & .0022 & .0093 & .0039 \\
\addlinespace[1pt]
$[11,15]$ & RPG & .0179 & .0110 & .0290 & .0146 & .0091 & .0052 & .0171 & .0078 \\
$[11,15]$ & SAHL & .0285 & .0205 & .0459 & .0261 & .0133 & .0078 & .0261 & .0119 \\
\addlinespace[1pt]
$[16,20]$ & RPG & .0159 & .0100 & .0286 & .0141 & .0191 & .0111 & .0315 & .0152 \\
$[16,20]$ & SAHL & .0244 & .0165 & .0433 & .0227 & .0224 & .0130 & .0410 & .0189 \\
\addlinespace[1pt]
$>20$ & RPG & .0666 & .0451 & .0949 & .0542 & .0893 & .0605 & .1288 & .0733 \\
$>20$ & SAHL & .0739 & .0517 & .1043 & .0614 & .1080 & .0740 & .1555 & .0893 \\
\bottomrule
\end{tabular*}
\endgroup
\end{minipage}\par\medskip

\par\medskip
\noindent\begin{minipage}{\linewidth}
\centering
\captionsetup{hypcap=false}
\captionof{table}{Detailed results for Figure~\ref{fig:sequence_length}: test-history-length groups.}
\label{tab:app_history_length}
\begingroup
\small
\renewcommand{\arraystretch}{1.08}
\setlength{\tabcolsep}{2.0pt}
\begin{tabular*}{\linewidth}{@{\extracolsep{\fill}}llrrrrrrrr@{}}
\toprule
\textbf{History length} & \textbf{Method} & \multicolumn{4}{c}{\textbf{Beauty}} & \multicolumn{4}{c}{\textbf{Toys and Games}} \\
 &  & R@5 & N@5 & R@10 & N@10 & R@5 & N@5 & R@10 & N@10 \\
\midrule
$\leq 10$ & RPG & .0451 & .0319 & .0664 & .0387 & .0545 & .0385 & .0773 & .0458 \\
$\leq 10$ & SAHL & .0564 & .0406 & .0781 & .0476 & .0711 & .0524 & .0967 & .0606 \\
\addlinespace[1pt]
$[11,20]$ & RPG & .0586 & .0397 & .0795 & .0465 & .0410 & .0266 & .0637 & .0339 \\
$[11,20]$ & SAHL & .0673 & .0494 & .0988 & .0597 & .0645 & .0454 & .0931 & .0547 \\
\addlinespace[1pt]
$[21,30]$ & RPG & .0779 & .0581 & .1084 & .0681 & .0266 & .0209 & .0615 & .0323 \\
$[21,30]$ & SAHL & .0840 & .0645 & .1298 & .0792 & .0635 & .0449 & .0799 & .0502 \\
\addlinespace[1pt]
$[31,40]$ & RPG & .0996 & .0661 & .1255 & .0745 & .0547 & .0391 & .0846 & .0489 \\
$[31,40]$ & SAHL & .1299 & .1021 & .1905 & .1218 & .0896 & .0581 & .1144 & .0663 \\
\addlinespace[1pt]
$[41,50]$ & RPG & .1237 & .0918 & .1340 & .0949 & .0857 & .0507 & .1429 & .0690 \\
$[41,50]$ & SAHL & .1340 & .1053 & .1546 & .1120 & .1000 & .0736 & .1571 & .0933 \\
\addlinespace[1pt]
$>50$ & RPG & .1307 & .1030 & .1503 & .1094 & .0932 & .0548 & .1017 & .0577 \\
$>50$ & SAHL & .1569 & .1065 & .1961 & .1190 & .0678 & .0408 & .1186 & .0575 \\
\bottomrule
\end{tabular*}
\par\medskip
\begin{tabular*}{\linewidth}{@{\extracolsep{\fill}}llrrrrrrrr@{}}
\toprule
\textbf{History length} & \textbf{Method} & \multicolumn{4}{c}{\textbf{Sports and Outdoors}} & \multicolumn{4}{c}{\textbf{CDs and Vinyl}} \\
 &  & R@5 & N@5 & R@10 & N@10 & R@5 & N@5 & R@10 & N@10 \\
\midrule
$\leq 10$ & RPG & .0291 & .0192 & .0432 & .0237 & .0480 & .0323 & .0697 & .0393 \\
$\leq 10$ & SAHL & .0369 & .0256 & .0554 & .0315 & .0587 & .0398 & .0863 & .0487 \\
\addlinespace[1pt]
$[11,20]$ & RPG & .0269 & .0186 & .0425 & .0236 & .0407 & .0274 & .0612 & .0340 \\
$[11,20]$ & SAHL & .0340 & .0231 & .0494 & .0281 & .0494 & .0330 & .0741 & .0409 \\
\addlinespace[1pt]
$[21,30]$ & RPG & .0221 & .0129 & .0291 & .0152 & .0399 & .0254 & .0596 & .0318 \\
$[21,30]$ & SAHL & .0245 & .0184 & .0420 & .0239 & .0518 & .0346 & .0786 & .0433 \\
\addlinespace[1pt]
$[31,40]$ & RPG & .0332 & .0239 & .0415 & .0264 & .0469 & .0316 & .0701 & .0391 \\
$[31,40]$ & SAHL & .0498 & .0385 & .0581 & .0412 & .0514 & .0362 & .0751 & .0439 \\
\addlinespace[1pt]
$[41,50]$ & RPG & .0112 & .0056 & .0112 & .0056 & .0400 & .0269 & .0589 & .0331 \\
$[41,50]$ & SAHL & .0112 & .0056 & .0112 & .0056 & .0410 & .0298 & .0629 & .0367 \\
\addlinespace[1pt]
$>50$ & RPG & .0449 & .0253 & .0449 & .0253 & .0308 & .0204 & .0471 & .0257 \\
$>50$ & SAHL & .0337 & .0175 & .0562 & .0248 & .0408 & .0278 & .0593 & .0338 \\
\bottomrule
\end{tabular*}
\endgroup
\end{minipage}\par\medskip

\end{document}